\documentclass{article}
\usepackage[a4paper,
            bindingoffset=0.2in,
            left=0.6in,
            right=0.6in,
            top=1in,
            bottom=1in,
            footskip=.25in]{geometry}
\usepackage[utf8]{inputenc}
\usepackage{jcappub}
\usepackage{natbib}
\setcitestyle{square,numbers}
\usepackage{graphicx}
\usepackage{xcolor}
\usepackage{float}
\usepackage{multirow,array,longtable}
\usepackage{arydshln} 
\usepackage[makeroom]{cancel}
\usepackage{amsfonts}

\renewcommand\eqref[1]{Eq.~(\ref{#1})}
\newcommand\eqrefs[2]{Eqs.~(\ref{#1})-(\ref{#2})}
\newcommand\figref[1]{Fig.~\ref{#1}}

\newcommand\tabref[1]{Table~\ref{#1}}

\newcommand\secref[1]{Section~\ref{#1}}
\newcommand\appref[1]{Appendix~\ref{#1}}
\newcommand{\be}{\begin{equation}}
\newcommand{\ee}{\end{equation}}
\newcommand{\bear}{\begin{eqnarray}}
\newcommand{\eear}{\end{eqnarray}}
\newcommand{\nn}{\nonumber}

\def\1loop{one-loop}
\def\amp{\mathcal{A}}
\newcommand{\mL}{\mathcal{L}}
\newcommand{\mO}{\mathcal{O}}

\def\lsm{\lambda}
\def\cw{c_{\rm w}}
\def\sw{s_{\rm w}}

\def\mh{m_{h}}
\def\mw{m_{W}}
\def\mz{m_{Z}}
\def\gY{g'}
\def\gYc{g^{'2}}

\def\vev{{\it v}}
\def\xh{\frac{4\mw^2}{\mh^2}}
\def\xz{\frac{4\mw^2}{\mz^2}}

\def\cb{c_\beta}
\def\sb{s_\beta}
\def\tanb{t_\beta}

\newcommand{\sba}{s_{\beta-\alpha}}
\newcommand{\cba}{c_{\beta-\alpha}}
\def\mHp{m_{H^\pm}}
\def\mHeavy{m_H}
\def\mA{m_A}
\def\lahHpHm{\lambda_{hH^+H^-}}
\def\lahhH{\lambda_{hhH}}
\def\lahhh{\lambda_{hhh}}

\def\lahhhh{\lambda_{hhhh}}

\def\mbarc{\frac{m_{12}^2}{s_\beta c_\beta}}
\def\mbarh{\frac{m_{12}^2}{\mh^2s_\beta c_\beta}}

\title{Non-decoupling effects from heavy Higgs bosons by matching 2HDM to HEFT amplitudes}
\author[a]{F. Arco,}
\author[a]{D. Domenech,}
\author[a]{M. J.  Herrero,}
\author[b]{and R. A.  Morales}
\affiliation[a]{Departamento de F\'{\i}sica Te\'orica and Instituto de F\'{\i}sica Te\'orica, IFT-UAM/CSIC,\\
Universidad Aut\'onoma de Madrid, Cantoblanco, 28049 Madrid, Spain}
\affiliation[b]{IFLP, CONICET - Dpto. de F\'{\i}sica, Universidad Nacional de La Plata, \\ 
C.C. 67, 1900 La Plata, Argentina}
\emailAdd{francisco.arco@uam.es}
\emailAdd{jose.domenech@estudiante.uam.es}
\emailAdd{maria.herrero@uam.es}
\emailAdd{roberto.morales@fisica.unlp.edu.ar}

\abstract{In this work we explore the low energy effects induced from the integration of the heavy Higgs boson modes,  $H$,  $A$ and $H^\pm$,  within the Two Higgs Doublet Model (2HDM) by assuming that the lightest Higgs boson $h$ is the one observed experimentally at $m_h \sim 125$ GeV.  We work within the context of Effective Field Theories,  focusing on the Higgs Effective Field Theory (HEFT),  although some comparisons with the Standard Model Effective Field Theory (SMEFT) case are also discussed through this work.  Our main focus is placed in the computation of the non-decoupling effects from the heavy Higgs bosons and the capture of such effects by means of the  HEFT coefficients which are expressed in terms of the input parameters of the 2HDM.  
Our approach to solve this issue is by matching the amplitudes of the 2HDM and the HEFT for physical processes involving the light Higgs boson $h$ in the external legs,  instead of the most frequently used matching procedure at the Lagrangian level.  More concretely, we perform the matching at the amplitudes level for the following physical processes,  including scattering and decays: $h\to WW^*\to Wf\bar{f'}$,  $h\to ZZ^*\to Zf\bar{f}$,  $WW \to hh$,  $ZZ \to hh$,  $hh \to hh$,  $h \to \gamma \gamma$ and $h \to \gamma Z$.  One important point of this work  is that the matching is required to happen at low energies compared to the heavy Higgs boson masses,  and these are heavier than the other particle masses.  The proper expansion for this heavy mass limit  is also defined here,  which provides the results for the non-decoupling effects presented in this work.  
We finally discuss the implications of the resulting effective coefficients,  and remark on the interesting correlations detected among them.}

\begin{document}
\begin{flushright}
   \today\\
	IFT-UAM/CSIC-23-97 
\end{flushright}
\maketitle
\section{Introduction}
\label{section-intro}

The use of Effective Field Theories (EFTs) to describe new physics effects from scenarios  beyond the Standard Model (BSM)
is nowadays a powerful and convenient tool in many aspects.  Firstly,  because they are built mainly from symmetry requirements,  therefore providing  a model independent framework.  Secondly,  because the absence of new particles discoveries at the experiments indicates no preference from data for any particular fundamental underlying theory to describe the BSM physics.  In the EFT context the information on the new physics is exclusively contained in the specific values of the effective coefficients (named in different ways in the literature: Wilson coefficients,  effective parameters,  effective low energy constants, etc.).  On the other hand,  the direct comparison with data of the predictions from EFTs for observable quantities  is a valuable task and of great interest nowadays.  In particular, we focus here on the EFTs that describe the BSM Higgs boson physics and that contain the Higgs boson observed experimentally with a mass value of $m_h \sim 125$ GeV~\cite{ATLAS:2012yve,CMS:2012qbp,ATLAS:2015yey}.  The two most popular EFTs are the SMEFT (Standard Model Effective Field Theory) and the HEFT (Higgs Effective Field Theory), see for instance the reviews~\cite{Brivio:2017vri,Dobado:2019fxe}.  In contrast to the so-called $\kappa$-framework,  which is usually preferred by the experimental community to constraint the BSM effective couplings (also called anomalous effective couplings),  these two gauge theories are well defined quantum EFTs which are built under the gauge symmetry guiding principle.  SMEFT and HEFT  preserve both the gauge symmetries of the SM,   i.e they are both built from $SU(3)_C\times SU(2)_L\times U(1)_Y$ gauge invariant effective operators and both are renormalizable theories,  using the more relaxed definition of renormalization in EFTs.  This accounts for renormalization in the truncated series of effective operators for the given EFT.

In this work,  we choose the HEFT to describe the BSM Higgs physics because it is better suited for scenarios where the underlying UV theory generating such EFT at low energies could be strongly interacting.  By strongly interacting dynamics here we simply mean UV scenarios that include the possibility of  large couplings,  in contrast to weakly interacting scenarios where all couplings are small.  We will in particular focus here on the bosonic part of the HEFT whose effective Lagrangian was called in its origins Electroweak  Chiral Lagrangian (EChL), in close analogy to the Chiral Lagrangian (ChL) and Chiral Perturbation Theory (ChPT) for low energy QCD.  Similarly to ChPT being the proper tool to describe the low energy hadronic physics with QCD as the UV theory containing large couplings,  the HEFT is the proper tool to describe the low energy Higgs physics from  the given UV theory containing large  couplings.  

Our main interest here is the use of  HEFT as the proper tool to capture all the potential non-decoupling effects from the heavy BSM Higgs bosons in the case where the UV underlying theory is the well known Two Higgs Doublet 
Model (2HDM), for a review see for instance~\cite{HHG,Branco:2011iw}.  Our single assumption for the 2HDM is the following:  out of the five physical Higgs bosons,  the lightest one $h$ is identified with the observed Higgs particle, and the other four $H$, $H^\pm$ and $A$ are assumed to be heavier than the EW scale,  namely, heavier than $ v=246\, {\rm GeV}$.   Given the present constraints from all experimental searches this seems a very reasonable assumption.  Then,  under this assumption  the immediate question comes,  what are the low energy effects from the integration out of these heavy modes (tree and one-loop level) that could be observed in an experiment?  In general, if the heavy modes leave non-decoupling effects in the low energy observables they could be more easily detected, in contrast to the so-called decoupling effects that leave weaker hints in the low energy physics.  To be more precise,  the definition of decoupling versus non-decoupling effects in the low energy physics is well established in the famous decoupling theorem of Appelquist and Carrazone~\cite{PhysRevD.11.2856}.  In short, it is an statement on the behavior with the heavy particle mass of the 1PI  proper vertices with light particles in the external legs after the heavy particles have been integrated out  (at any order).  It says that there is decoupling when all these heavy particle effects can be absorbed into redefinitions of the couplings, parameters  and fields of the low energy theory (i.e.  renormalizing these quantities) or else they are suppressed by inverse powers of the heavy masses.  In contrast,  when this behavior does not happen,  and the heavy particle  effects in the physical observables do not decrease as inverse powers of the heavy masses (hence leading to hints at low energies) they are said to be non-decoupling.  The most clear examples of non-decoupling effects appear in theories with spontaneous symmetry breaking and,  in particular,  when the particle masses are generated by a Higgs mechanism,  providing a relation between the generated mass, the coupling and the vacuum expectation value which defines the broken phase.  For instance, within the SM itself it applies to the Higgs boson case with self-coupling and mass being related by $\lambda =m_h^2/(2v^2)$.  It also happens in the top quark case with Yukawa coupling and mass being related by $y_t=\sqrt{2}m_t/v$.  These two particles leave non-decoupling effects (i.e.  non decreasing with inverse powers of their masses) in several observables at low energies which indeed have been explored in the past experiments.  For instance in $\Delta\rho$ which defines the radiative corrections to the $W$ and $Z$ boson mass relation,  $\rho=m_W^2/(m_Z^2 \cos^2 \theta_W)$,  with respect to the tree-level prediction, $\rho^{\rm tree}=1$.  
These and other non-decoupling effects  from the heavy SM Higgs in the 1PI one-loop functions with external EW gauge bosons were computed long ago and collected in a set of effective coefficients of the EChL in~\cite{Herrero:1993nc,Herrero:1994iu}.  Similarly,  we aim to explore here the 2HDM case with the BSM heavy Higgs bosons leaving non-decoupling effects in low energy observables which can be collected in a set of effective coefficients of the HEFT.

It should be noticed,  that these kind of BSM non-decoupling effects cannot be encoded within the SMEFT framework  since,  by construction,  the effective operators describing the BSM Higgs physics at low energies carry Wilson coefficients that are suppressed by inverse powers of the UV energy scale.  Therefore, in the present case of integrating out the 2HDM heavy Higgs bosons they lead to decoupling effects that go as inverse powers of the heavy Higgs boson masses $m_{\rm heavy}$ , i.e.  $\sim (1/m_{\rm heavy}^2)$ for dimension 6 operators,  $\sim (1/m_{\rm heavy}^4)$ for dimension 8 operators, and so on. 
For instance,  in~\cite{Belusca-Maito:2016dqe},  the SMEFT dimension six low energy effects for 2HDM are derived at the Lagrangian level and they find such decoupling behavior with the heavy mass.
The necessity to move
beyond dimension six interactions within the SMEFT for any scenario that contains Higgs boson mixing and the inclusion of dimension eight operator effects are discussed in~\cite{Dawson:2022cmu,Banerjee:2022thk,Banerjee:2023bzl}, where they also find such decoupling behavior. 
Then,  within the SMEFT,  these decoupling effects from the heavy Higgs bosons disappear, in the heavy mass limit, and the 2HDM converges to the SM.  The case of the HEFT is different,   and the convergence of the 2HDM to the SM is reached in a different way,  as will be discussed here.   The relation between the HEFT and the SMEFT is by itself an interesting field of research and it has been considered under different approaches.  For instance,  a geometrical approach was considered in~\cite{Alonso:2015fsp,Alonso:2016oah,Cohen:2020xca}.  Some relations between the SMEFT and the HEFT, within the context of the 2HDM,  have also been considered recently in~\cite{Dawson:2023ebe}.  They require matching at the Lagrangian level and also require decoupling and perturbativity as a principle guide,  arriving to results for the amplitudes were they find no relevant differences among  HEFT and SMEFT.  In summary,   all effects from the 2HDM heavy Higgs bosons being described by EFTs  in the previous works are found to be decoupling at low energy observables.  This is in contrast with our study here. 

In this work we will present a computation of the non-decoupling effects from the heavy Higgs bosons of the 2HDM by matching  the predictions at the amplitude level of the 2HDM with the predictions from the HEFT, considering the leading effects in each observable,  either at the tree level or the one-loop level depending on the process. The result of this matching will provide the values of the HEFT effective coefficients containing these non-decoupling effects.  It should be noticed that,  {\it a priori},  a non-decoupling behavior is expected to happen in the 2HDM case  because the triple Higgs couplings can have large values due to the relations of these couplings with the heavy masses.  For instance,  $\lambda_{hH^+H^-}$ can be large for heavy $m_{H^\pm}$ since the derived $\lambda_{hH^+H^-}$ in terms of the physical masses contains a ${\cal O}(m_{H^\pm}^2/v^2)$ term.  One crucial difference with respect to other approaches is that we have chosen here to do this matching of predictions at the amplitude level, i.e with observable/measurable physical quantities.  In general,  there are three alternatives to do matching among the UV theory and the low energy EFT, and they are not totally equivalent.  The matching can be done: 1) at the Lagrangian level,  2) at the effective action level (or, equivalently,  identifying  the full set of 1PI functions) and 3) at the amplitude level.  The simplest and  most frequently used method in the literature is the first one.  The most complete framework for matching  is the second one, since it implies the identification in the two theories of all the 1PI renormalized functions with external light particles, being generically off-shell.   However,  we have preferred to match amplitudes (with external physical particles on-shell)  since we believe it is more physical,  free from ambiguities in field redefinitions, choice of operator basis and renormalization prescriptions.  Furthermore,  it does not require the use of pseudo-observables (like in the $\kappa$ framework) to connect with data since the prediction of the amplitude  is directly comparable with data.  By  requiring the matching at the amplitude level between the predictions from the HEFT and the 2HDM with large heavy Higgs masses, and solving these matching equations we will be able to extract the values of the HEFT coefficients in terms of the 2HDM input parameters. These are chosen here to be the physical masses,  $m_h$, $m_H$, 
$m_{H^\pm}$,  $m_A$,  the ratio of the two Higgs vev's,  $\tan\beta$,  the parameter $\cos(\beta-\alpha)$ and the $Z_2$ soft-breaking mass parameter $m_{12}$.   To be more precise we do the matching of the 2HDM and HEFT amplitudes after performing a large mass expansion of the 2HDM amplitude in terms of the inverse powers of the heavy Higgs boson masses.   Notice that this is a well defined and convergent expansion,  as it will be shown here.  We have selected to match the amplitudes of some specific processes (scattering and decays)  involving the light Higgs boson in the external legs  which contain the most relevant non-decoupling effects.  Concretely:  $h\to WW^*\to Wf\bar{f'}$,  $h\to ZZ^*\to Zf\bar{f}$,  $WW \to hh$,  $ZZ \to hh$,  $hh \to hh$,  $h \to \gamma \gamma$ and $h \to \gamma Z$.   The explicit computations of the 2HDM amplitudes for these processes will be presented and discussed here.  Some of these processes have also been considered within the HEFT previously in the literature for different purposes.   In particular,  $WW \to hh$,    $h \to \gamma \gamma$ and $h \to \gamma Z$ have been computed within the NLO-HEFT including renormalization of the one-loop corrections and doing the computation in a generic $R_\xi$ gauge~\cite{paperHdecays,Herrero:2021iqt, Herrero:2022krh}.  The cases of $WW \to hh$ and $hh \to hh$ scattering processes have also been studied within the NLO-HEFT in~\cite{Delgado:2013hxa, Delgado:2014dxa, Dobado:2019fxe, Asiain:2021lch, Asiain:2023zhx}.  The Higgs decays  
$h\to WW^*\to Wf\bar{f'}$,  $h\to ZZ^*\to Zf\bar{f}$,  $h \to \gamma \gamma$,  $h \to \gamma Z$ and the Vector Boson Fusion scattering $VV \to hh$  have also been studied within the NLO-HEFT in~\cite{Anisha:2022ctm} focusing in the  implications for LHC physics.  In the present paper,  it will be shown  that in order to capture the leading non-decoupling effects from the 2HDM heavy Higgs bosons in  these processes it is sufficient to do the matching at ${\cal O} (\hbar^0)$ in all cases except in the $h \to \gamma \gamma$ and $h \to \gamma Z$ decays where the matching must be done at ${\cal O}(\hbar^1)$.  Once we solve analytically these matching equations and find the expressions of the effective coefficients in terms of the 2HDM input parameters,  we will  analyze the predictions for those coefficients,  both analytically and numerically.  We will do that analysis in several interesting scenarios that, following the usual terminology,  we classify as: 1) alignment ($\cos (\beta-\alpha) = 0$),  2) misalignment  ($\cos (\beta-\alpha) \neq 0$),  and 3) quasi-alignment  ($\cos (\beta-\alpha)\ll 1$).  In the final part of this work we will discuss on the interesting correlations found here among the HEFT coefficients in these three  different limits.  

The paper is organized as follows: \secref{section-EChL} provides a brief introduction to the HEFT focusing on its comparison to the SMEFT approach. The relevant part of the HEFT Lagrangian for the present computation is also included in this section.  \secref{section-2HDM} contains the relevant details of the 2HDM, in particular,  the expressions of the derived triple Higgs couplings in terms of the selected input parameters of this model. The analytical expressions of the amplitudes for the considered processes in the SM, HEFT and 2HDM are collected in \secref{section-amplitudes}. The matching procedure is described in \secref{section-matching}.  The way to the solution for the HEFT coefficients in terms of the input 2HDM parameters that summarizes  the non-decoupling effects in the heavy mass limit is also included in this section.  Also, the correlations found among the HEFT coefficients from the 2HDM are presented  in this section. The numerical analysis of the previous results are presented in \secref{section-numerical}. Finally, we conclude in \secref{section-conclus}. The details of the Feynman rules and the one-loop functions are given in the appendices.

\section{HEFT versus SMEFT}
\label{section-EChL}
As stated above, we have focused in this paper on the HEFT.  However,  before reviewing the details of the HEFT Lagrangian that are needed for the present computation,  we would like first to comment shortly on the most relevant aspects that crucially differentiate HEFT versus SMEFT. These differences will allow us to better understand the decoupling versus the non-decoupling effects of the heavy Higgs boson modes in the two low energy theories. 

The  main differences between these two EFTs can be summarized as  follows.  1) In the SMEFT the Higgs boson is introduced as a component of a $SU(2)$ doublet whereas in the HEFT it is introduced as a singlet. 2) The would-be Goldstone bosons (GBs) associated to the electroweak  (EW) symmetry breaking,  $SU(2)_L\times U(1)_Y \to U(1)_{\rm em}$,  in the SMEFT are identified with the other components in this doublet,  therefore completing together the simplest  linear realization.  In contrast,  in the HEFT the GBs  transform non-linearly under the  global symmetry of the scalar sector $SU(2)_L \times SU(2)_R$,    usually called the EW chiral symmetry,  and they are frequently parametrized by an exponential function. 3) The ordering of the effective operators in the SMEFT is done in terms of the canonical dimension (cd = 4, 6, 8 etc),  whereas in the HEFT this ordering is done in terms of the chiral dimension (chd=2,4 etc).  This different counting (cd versus chd)  leads to a different classification in both theories of what is the leading order (LO) versus what is the next to leading order (NLO).  The ordering in cd implies that the LO SMEFT Lagrangian with cd= 4 is the SM Lagrangian,  the NLO SMEFT includes the cd=6 operators with coefficients being suppressed by inverse powers of the ultraviolet (UV) $ \Lambda$ scale,  as  $\sim {\Lambda}^{-2}$, and so on.  In contrast,  the  HEFT reaches the SM for an specific choice of the effective coefficients in the LO Lagrangian with chd=2 and setting  the NLO effective coefficients to zero.  4) The predictions for observables in both EFTs are also very different.   In particular the HEFT,  due to the chd counting (involving derivatives and soft masses),  provides predictions for the amplitudes of physical processes  that are organized typically as expansions in powers of the relevant process energy (and powers of the soft masses involved).  This is not the case in the SMEFT,  which in contrast provides predictions for the amplitudes that are organized in terms of the expansion in inverse powers of  $\Lambda$.  5) The renormalization programs in both EFTs are also very different.  In the SMEFT,  all the operators are renormalized together,  without doing any distinction in the renormalization procedure between the LO and  the NLO contributions.  In contrast,  in the HEFT there is a hierarchy in the renormalization procedure between the LO and the NLO contributions.  In the chiral counting the divergent loops computed with the LO  Lagrangian  (chd=2) are renormalized by the effective coefficients of the NLO  Lagrangian (chd=4),  providing a well defined framework for renormalization to one-loop order (like in ChPT),  with a marked hierarchy LO/NLO and where the relevant scale in this loop counting is given typically by $4 \pi v \sim 3\,  {\rm TeV}$.   The 
one-loop renormalization program in the bosonic sector of the HEFT has been studied using a $R_\xi$ general covariant gauge in~\cite{paperHdecays,Herrero:2021iqt,Herrero:2022krh}.   The renormalization in the SMEFT,  via the renormalization group equations,   was studied in~\cite{Jenkins:2013zja,Jenkins:2013wua,Alonso:2013hga}.
The 
SMEFT for $R_\xi$ gauges was studied in~\cite{Dedes:2017zog, Misiak:2018gvl}.
Some illustrative discussion on the matching at the 1PI functions level can also be found in~\cite{Brivio:2021alv}. 
On the other hand,   the matching of tree-level amplitudes predicted by the HEFT and the SMEFT also leads to some relations between the HEFT and the SMEFT coefficients~\cite{RoberMariaDaniMJ}.  But all these attempts of matchings  among the HEFT and the SMEFT provide just partial relations among these two theories since they assume a specific  given order in the expansion of both theories (LO, NLO etc) and also a given order in the loop expansion (i.e in the ${\cal O}(\hbar^n)$ expansion),  so they are not complete comparisons.  Furthermore, to conclude that one theory contains the other one should proceed with a complete comparison of the full quantum EFTs (i.e not truncated, and to all orders in the loop expansion).  But this is a difficult task.

Next we proceed with the short summary of the needed ingredients of the HEFT Lagrangian for the present computation.  For this presentation and the notation we follow closely~\cite{paperHdecays,Herrero:2021iqt,Herrero:2022krh}.  First, we recall that  in the bosonic sector the active degrees of freedom  are:  the EW gauge bosons, $B_\mu$ and $W^a_\mu$ ($a=1,2,3$), their corresponding GBs $\pi^a$ ($a=1,2,3$), and the Higgs boson $h$. The Lagrangian is invariant under EW gauge,  $SU(2)_L\times U(1)_Y$,  transformations and the scalar sector of the EChL has an additional invariance under the EW chiral $SU(2)_L \times SU(2)_R$ symmetry.
The Higgs boson field is invariant under all transformations, i.e., it is a singlet of the EW chiral symmetry and the EW gauge symmetry. Therefore the interactions of $h$  are introduced via generic polynomials since there are not limitations from symmetry arguments.
On the other hand, the GBs $\pi^a$ ($a=1,2,3$) transform non-linearly under this EW chiral transformations. Then they are introduced in a non-linear representation via the exponential parametrization, by means of the matrix $U$, which transforms linearly under the EW chiral transformations:
\be 
U(\pi^a) = e^{i \pi^a \tau^a/\vev} \, \, , 
\label{expo}
\ee
where $\tau^a$, $a=1,2,3$,  are the Pauli matrices and $v=246$ GeV.  
In addition, the EW gauge bosons are introduced by the gauge invariance principle and they appear in the following combinations: 
\bear
\hat{B}_\mu &=& \gY B_\mu \tau^3/2\,, \quad \hat{W}_\mu = g W^a_\mu \tau^a/2 \,,\quad D_\mu U = \partial_\mu U + i\hat{W}_\mu U - i U\hat{B}_\mu \,,   \nn\\
\hat{B}_{\mu\nu} &=& \partial_\mu \hat{B}_\nu -\partial_\nu \hat{B}_\mu \,, \quad \hat{W}_{\mu\nu} = \partial_\mu \hat{W}_\nu - \partial_\nu \hat{W}_\mu + i  [\hat{W}_\mu,\hat{W}_\nu ] \,. 
\eear

In the chiral dimension counting, all derivatives and masses count as momentum: 
\be\partial_\mu \,,\,\mw \,,\,\mz \,,\,\mh \,,\,g\vev \,,\,\gY\vev \,,\lambda v\, \sim \mO(p)\, . 
\ee
The HEFT organizes the effective operators in the EChL into terms with increasing chiral dimension, starting at chd=2, then chd=4,  and so on. 
Notice again that this chiral counting differs from the usual SMEFT expansion with operators having growing canonical dimension,  starting at cd=4,  then cd=6 etc,  and  terms suppressed with the heavy scale $\Lambda$.

For  the bosonic sector of the HEFT,  we consider the leading order Lagrangian,  with chiral dimension two, $\mL_2$, and the next to leading order one with chiral dimension four, $\mL_4$:
\be
{\cal L}_{\rm HEFT}={\cal L}_{\rm EChL}=\mL_2+ \mL_4 \,,
\label{EChLorder}
\ee

Firstly, the leading order Lagrangian is given by,
\bear
\mL_2 &=& \frac{\vev^2}{4}\left(1+2a\frac{h}{\vev}+b\left(\frac{h}{\vev}\right)^2 +\ldots\right){\rm Tr}\Big[ 
 D_\mu U^\dagger D^\mu U \Big]+\frac{1}{2}\partial_\mu h\partial^\mu h-V_{\rm EChL}(h)  \nn\\
&&-\frac{1}{2g^2} {\rm Tr}\Big[ \hat{W}_{\mu\nu}\hat{W}^{\mu\nu}\Big]
-\frac{1}{2\gYc}{\rm Tr}\Big[ 
\hat{B}_{\mu\nu}\hat{B}^{\mu\nu}\Big] +\mL_{GF} +\mL_{FP}  \,.
\label{eq-L2}
\eear
Here $V_{\rm EChL}(h)$ is the EChL Higgs potential, $\mL_{GF}$ and $\mL_{FP}$,  are the gauge-fixing and Faddeev–Popov Lagrangian, respectively.
The dots stand for terms that do not enter in our processes of interest, neither at tree level nor at \1loop level. 
The EChL Higgs potential in $\mL_2$  is given by:
\be
V_{\rm EChL}(h) = \frac{\mh^2}{2}h^2 +\kappa_3\lambda\vev\, h^3+\kappa_4\frac{\lambda}{4}h^4 \,,
\label{EChL_Higgs-pot}
\ee
where $\mh^2=2\lambda\vev^2$ as in the SM, and values for $\kappa_3$ and $\kappa_4$ different from 1 encode the physics beyond SM.

We implement the linear covariant $R_\xi$ gauges~\cite{Fujikawa:1972fe} with the gauge-fixing Lagrangian given by, 
\bear
\mL_{\rm GF} &=& -F_+F_- -\frac{1}{2}F_{Z}^2 -\frac{1}{2}F_{ A}^2  \nn\\
&=& -\frac{1}{\xi}(\partial^{\mu}W_{\mu}^+ -\xi\mw \pi^+)(\partial^{\mu}W_{\mu}^- -\xi\mw \pi^-) -\frac{1}{2\xi}(\partial^{\mu}Z_{\mu}-\xi \mz \pi^{3})^2 -\frac{1}{2\xi}(\partial^{\mu}A_{\mu})^2 \,,
\label{GF-lag}
\eear
and the corresponding Faddeev-Popov Lagrangian~\cite{Faddeev:1967fc},  given by:
\be
\mL_{\rm FP} = \sum_{i,j=+,-,Z,A} \bar{c}^{i} \frac{\delta F_i}{\delta \alpha_j} c^j \,,
\label{FP-lag}
\ee
where $\xi$ is the generic gauge-fixing parameter of the $R_\xi$ gauges, $c^j$ are the ghost fields and $\alpha_j$ ($j=+,-, Z,A$) are the corresponding gauge transformation parameters.
Notice that $\mL_{\rm GF}$ of \eqref{GF-lag} is the same as in the SM and for the 2HDM. Formally, the expression in \eqref{FP-lag} are also the same as in the SM and 2HDM.
However, the Higgs and GBs transformations in this non-linear EFT differ from the corresponding ones in the SM yielding to different interactions among those scalars and the ghost fields.

In the case that a given observable require a \1loop computation with the $\mL_2$ terms, the $\mL_4$ operators must be included in order to be consistent with the chiral counting and also to use the coefficients in $\mL_4$  as counterterms to renormalize the divergences generated by the loops from $\mL_2$,  following the usual  procedure with Chiral Lagrangians.  For the present work, we will compute the \1loop Higgs decays $h\to\gamma\gamma$ and $h\to\gamma Z$, then the relevant  terms of the next to leading order contributions  are included in the following Lagrangian~\cite{paperHdecays,Herrero:2021iqt,Herrero:2022krh}:
\bear
\mL_4 &=& - \left(a_{HBB} \frac{h}{\vev}+a_{HHBB} \frac{h^2}{\vev^2}\right) {\rm Tr} \Big[\hat{B}_{\mu\nu} \hat{B}^{\mu\nu} \Big]  - \left(a_{HWW} \frac{h}{\vev}+a_{HHWW} \frac{h^2}{\vev^2}\right) {\rm Tr}\Big[\hat{W}_{\mu\nu} \hat{W}^{\mu\nu}\Big]  \nn\\
&&+ \left(a_{H1} \frac{h}{\vev}+a_{HH1} \frac{h^2}{\vev^2}\right) {\rm Tr}\Big[ U \hat{B}_{\mu\nu} U^\dagger \hat{W}^{\mu\nu}\Big]  + \ldots 
\label{eq-L4relevant}
\eear
where the relevant coefficients for the Higgs decays under consideration,  $h\to\gamma\gamma$ and $h\to\gamma Z$,  are given in terms of the coefficients in \eqref{eq-L4relevant}  by the following relations:
\bear
a_{h\gamma\gamma} &=& a_{HBB}+a_{HWW}-a_{H1} \,,  \\
a_{h\gamma Z} &=& \frac{1}{\cw^2}(-a_{HBB} \sw^2+a_{HWW}\cw^2-\frac{1}{2} a_{H1} (\cw^2-\sw^2)) \,.
\label{HM1}
\eear
These coefficients in \eqref{eq-L4relevant} and others (see~\cite{paperHdecays,Herrero:2021iqt,Herrero:2022krh}) also enter in other observables when they are predicted at the one-loop level.  In addition to the mentioned scattering amplitudes $WW \to hh$ and $ZZ \to hh$,  these coefficients will also enter in other vector boson scattering amplitudes for double Higgs production  when computed to one-loop like $ \gamma \gamma \to hh$ and $\gamma Z \to hh$.  Nevertheless, we do not study these one-loop amplitudes here,  since the focus of our interest is to capture the most relevant non-decoupling effects from the heavy modes of the UV theory,  which, as we will see, are summarized in the tree-level coefficients from $\mL_2$,  $a$, $b$, $\kappa_3$, $\kappa_4$,   and in the NLO coefficients from $\mL_4$,  
$a_{h\gamma\gamma}$ and $a_{h\gamma Z}$.  

One important point to recall is the way the SM is embedded within the HEFT.  It is clear that the comparison cannot be done at the Lagrangian level,  since within the SM the Higgs field is introduced into a doublet whereas in the HEFT it is a singlet.  Thus, the Lagrangian themselves are not directly comparable.  The Feynman rules for the couplings with scalar fields ($h$ and the GBs)   are also different in the HEFT and the SM due to the non-linear parametrization used in the HEFT.  Thus,  the comparison between the SM and the HEFT should be done via their predictions for the observables instead of via their corresponding Lagrangians.  Specifically,  in order to reach the SM predictions from the HEFT predictions one has to fix the HEFT coefficients as follows: 1) the LO coefficients must be set to the following particular values: $a=1$,  $b=1$, $\kappa_3=1$ and $\kappa_4=1$.  Equivalently,  if we write the coefficients in terms of the corresponding $\Delta$'s,  which are  defined by $a=1-\Delta a$,  $b=1-\Delta b$, $\kappa_3= 1-\Delta \kappa_3$,   $\kappa_4= 1-\Delta \kappa_4$,  then the SM predictions can be obtained from the HEFT predictions by setting,  $\Delta a =0$,  $\Delta b =0$,  $\Delta \kappa_3=0$,  $\Delta \kappa_4=0$.  Additionally,  all the NLO coefficients must also be set to zero, i.e , generically $a_i=0$ for all coefficients in $\mL_4$.

We have summarized  the Feynman rules of the HEFT that are relevant for the present computation in \tabref{relevant-FR} of \appref{App-Frules} (for a full set see \cite{paperHdecays,Herrero:2021iqt,Herrero:2022krh}). The corresponding Feynman rules of the SM are also included for comparison. 

The embedding of the SM into the SMEFT is very different than in the HEFT, since the SM Lagrangian is explicitly included in the SMEFT Lagrangian as its first term contribution of canonical dimension 4:  

\begin{equation}
    \mathcal{L}_{\rm SMEFT} = \mathcal{L}_{\rm SM} + \mathcal{L}_6 + \mathcal{L}_8 + \dots\,,~~~\text{with}~\mathcal{L}_d = \frac{c_i}{\Lambda^{d-4}}  \mathcal{O}_i^{(d)}\,.
\end{equation}
And both the SM and the SMEFT place the Higgs boson into the standard Higgs doublet.  It is usually parametrized as follows:
\be
\Phi =\binom{-i\pi^+}{\frac{\vev+h-i\pi^3}{\sqrt{2}}} = \binom{G^+}{\frac{\vev+h+iG^0}{\sqrt{2}}} .
\label{eq:doublet}
\ee

Thus, to reach the SM predictions from the SMEFT predictions, this can be done at the Lagrangian level, by  simply setting all Wilson coefficients  $c_i$ to zero in the Lagrangian terms with canonical dimension 6, 8 etc. 

Finally,  before ending this section, it is worth recalling the previous relations found in~\cite{RoberMariaDaniMJ}  among the HEFT and SMEFT coefficients by the same procedure that we choose in the present paper of matching amplitudes.   This matching was done for the particular scattering process $WW \to hh$,  and the assumption for both EFTs was to work at the tree level and with the truncated Lagrangian,  $\mL_2+\mL_4$ for the HEFT and the truncated Lagrangian up to cd=8 for the SMEFT.  The result of this matching provides interesting relations among the effective coefficients of both theories.  These include the following relations (for a full set see~\cite{RoberMariaDaniMJ}):

\bear
 \Delta a \vert_{\rm SMEFT}&=&  - \frac{1}{4} \frac{\vev^2 }{\Lambda^2} \delta c_{\phi D}\,,   \nn\\
\Delta b  \vert_{\rm SMEFT}&=&  - \frac{\vev^2 }{\Lambda^2} \delta c_{\phi D} \,,  \nn\\
\Delta \kappa_3  \vert_{\rm SMEFT}&=- & \frac{5}{4} \frac{\vev^2 }{\Lambda^2} \delta c_{\phi D} \,, \nn\\
a_{HWW}  \vert_{\rm SMEFT}&=& -\frac{\vev^2}{2 \mw^2} \frac{\vev^2 }{\Lambda^2} c_{\phi W} \,, \nn\\
a_{HHWW}  \vert_{\rm SMEFT}&=& - \frac{\vev^2}{4\mw^2}\frac{\vev^2 }{\Lambda^2} c_{\phi W}  \,,
\label{eq:matchingSMEFT}
\eear
where the definitions for the Wilson coefficients $c_i$  above can also be found in the mentioned reference (there a different notation $a_i$ was used instead of the $c_i$ here).   
It is interesting to note that the above relations among the coefficients of the HEFT and the SMEFT occur across the different orders in both EFTs.  In particular,  LO coefficients of chd=2 in the HEFT appear related to coefficients of cd=6 in the SMEFT,  NLO coefficients of chd=4 in the HEFT are related to coefficients of cd=6 in the SMEFT (and also to the coefficients of cd=8) and so on.  This also implies that capturing the non-decoupling effects using the HEFT,  i.e.  effects non-suppressed by inverse powers of the heavy mass of the UV underlying theory,   cannot be reproduced by the SMEFT, since by construction all the UV effects in the SMEFT are suppressed by inverse powers of the heavy scale,  i.e.  they produce contributions in the amplitudes of ${\cal O}(1/\Lambda^2)$,   ${\cal O}(1/\Lambda^4)$, etc,  and they all decouple for large $\Lambda$.  Finally,  it is  worth mentioning that the previous values in \eqref{eq:matchingSMEFT} also indicates the existence of correlations among HEFT coefficients when they are matched to the SMEFT.  In particular,  in the subset given above, they are correlated as: $\Delta b  \vert_{\rm SMEFT}= 4 \Delta a \vert_{\rm SMEFT}$,  and $a_{HWW}  \vert_{\rm SMEFT} = 2 a_{HHWW}  \vert_{\rm SMEFT}$.

\section{Heavy Higgs bosons within the 2HDM}
\label{section-2HDM}
In this section we recall the basic aspects of the 2HDM that are relevant for the present computation.  
The 2HDM is the simplest extension of the SM  that includes two Higgs doublets,  $\Phi_1$ and $\Phi_2$,  instead of one doublet $\Phi$.  These two doublets are linear parametrizations of the four complex scalar fields (hence,  eight real scalar fields) defining the 2HDM  scalar sector.  They are usually defined as:
\begin{eqnarray}
\Phi_1 = \left( \begin{array}{c} \phi_1^+ \\ \frac{1}{\sqrt{2}} (v_1 +
    \rho_1 + i \eta_1) \end{array} \right) \;, \quad
\Phi_2 = \left( \begin{array}{c} \phi_2^+ \\ \frac{1}{\sqrt{2}} (v_2 +
    \rho_2 + i \eta_2) \end{array} \right) \;,
\label{eq:2hdmvevs}
\end{eqnarray}
where $v_1, v_2$ are the real vevs acquired by the fields
$\Phi_1, \Phi_2$, respectively, with $\tan\beta= v_2/v_1$ and they satisfy the 
relation $v = \sqrt{(v_1^2 +v_2^2)}$ where $v =246 \,  {\rm GeV}$ is the SM vev.
The eight degrees of freedom above, $\phi_{1,2}^\pm$, $\rho_{1,2}$ and
$\eta_{1,2}$, give rise to three Goldstone bosons, $G^\pm$ and $G^0$,
and five massive physical scalar fields: two $CP$-even scalar fields,
$h$ and $H$, one $CP$-odd one, $A$, and one charged pair, $H^\pm$.
Here the mixing angles $\alpha$ and $\beta$ diagonalize the $CP$-even and -odd
sectors,  respectively.  These rotations define the physical mass eigenstates,  $h$, $H$,  $A$ and $H^\pm$ in terms of the EW interaction eigenstates (or the other way around) and are given by:
\begin{eqnarray}
\phi_1^\pm&=&\cos \beta \, G^\pm -\sin \beta \, H^\pm,\nonumber \\
\phi_2^\pm&=&\sin \beta \, G^\pm +\cos \beta \, H^\pm, \nonumber \\
\eta_1&=&\cos \beta \, G^0 -\sin \beta \, A, \nonumber \\
\eta_2&=&\sin \beta \, G^0 +\cos \beta \, A, \nonumber \\ 
\rho_1&=&\cos \alpha \, H -\sin \alpha \, h, \nonumber \\
\rho_2&=&\sin \alpha \, H +\cos \alpha \, h. \nonumber
\end{eqnarray}
The relations among the two usual notations for the GBs inside the doublets are as in \eqref{eq:doublet}, i.e.  $G^{\pm}=- i \pi^{\pm}$,  $G^0=- \pi^0$. 

The self-interactions among the above scalar fields are provided by the 2HDM potential.  Since a general potential with two Higgs doublets can lead to flavor-changing neutral currents (FCNC) at the tree level, which are strongly discouraged by experimental measurements, we will  impose a $Z_2$ symmetry~\cite{Aoki:2009ha,Glashow:1976nt} meaning invariance under $\Phi_1\to\Phi_1$ and $\Phi_2\to-\Phi_2$.   Furthermore,  we will allow 
this $Z_2$ symmetry to be only softly broken by the parameter $m_{12}^2$, which has dimensions of mass squared. 
Thus, the relevant potential for the present work of the $CP$ conserving 2HDM with the $Z_2$ soft-breaking included,  expressed in terms of the two doublets $\Phi_1$ and $\Phi_2$,  is given by~\cite{HHG,Aoki:2009ha,Branco:2011iw}:
\begin{equation}
\begin{split}
V_{\rm 2HDM}(\Phi_1,\Phi_2) = m_{11}^2 (\Phi_1^\dagger\Phi_1) + m_{22}^2 (\Phi_2^\dagger\Phi_2) - m_{12}^2 (\Phi_1^\dagger
\Phi_2 + \Phi_2^\dagger\Phi_1) + \frac{\lambda_1}{2} (\Phi_1^\dagger \Phi_1)^2 +
\frac{\lambda_2}{2} (\Phi_2^\dagger \Phi_2)^2 \\
 + \lambda_3
(\Phi_1^\dagger \Phi_1) (\Phi_2^\dagger \Phi_2) + \lambda_4
(\Phi_1^\dagger \Phi_2) (\Phi_2^\dagger \Phi_1) + \frac{\lambda_5}{2}
[(\Phi_1^\dagger \Phi_2)^2 +(\Phi_2^\dagger \Phi_1)^2] .
\end{split}
\label{2HDM_Higgs-pot}
\end{equation}
After the EW symmetry breaking,  $SU(2)_L \times U(1)_Y \to U(1)_{\rm em}$,  the minimization conditions for the above 2HDM potential lead to the existence of five physical Higgs bosons: two $CP$-even Higgs bosons $h$ and $H$,  one $CP$-odd Higgs boson $A$ and  two charged Higgs bosons $H^\pm$ with masses given by $m_h$,  $m_H$ (with $m_h < m_H$),  $m_A$ and $m_{H^\pm}$ respectively.
In addition,  the three would-be Goldstone bosons disappear from the physical spectrum and provide the needed physical masses for the EW gauge bosons,  $m_W$ and $m_Z$. 
In this work, we will identify the $h$ state with the Higgs boson discovered in the LHC with a mass $m_h= 125$ GeV~\cite{ATLAS:2012yve,CMS:2012qbp,ATLAS:2015yey}.  The other Higgs bosons will be assumed here to be heavier than the EW scale $v$,  an hypothesis which is well justified given the present tight experimental constraints~\cite{Arco:2020ucn,Arco:2022xum}. 

The previous potential also contains the self-interactions among the scalars of the 2HDM which are very relevant for the present work.  In addition,  the interactions of the Higgs bosons with the gauge bosons are given by the gauge invariant Lagrangian built with the covariant derivatives of the two doublets and the interactions with fermions are given by the Yukawa part of the 2HDM Lagrangian.  They are well known in the literature and  we do not explicit them here for shortness (see for instance~\cite{HHG,Aoki:2009ha,Branco:2011iw}). 
The set of Feynman rules within the 2HDM that are relevant for the present computation are summarized in \tabref{relevant-FR} of \appref{App-Frules}. 

In order to make predictions for observables, one may use different choices for the 2HDM input parameters.  Here,  since we are going to consider later the hypothesis of very heavy BSM Higgs bosons,  and to deal with the expansion at large heavy masses, we believe that the most convenient choice for the input parameters should contain the physical Higgs masses.   Concretely we choose in this work the following 2HDM input parameters:
\begin{equation}
	v, \;\;\mh,\;\;\mHeavy,\;\;\mA,\;\;\mHp,\;\;\tanb,\;\;\cba,\;\;m_{12},
\label{eq-inputs}
\end{equation}
where we have adopted the shorthand notation $\cos x=c_x$, $\sin x=s_x$ and $\tan x=t_x$.  All the remaining 2HDM parameters and couplings are therefore derived quantities.

The choice of $\cba$ as an input parameter is motivated by the so-called \textit{alignment limit}, defined as ${\cba=0}$. 
Under this limit,  the interactions of $h$ with the gauge bosons and with the fermions recover their SM values.
For example,  the $h$ ($H$) coupling to $WW$ and $ZZ$ relative to the SM is given by $\sba$ ($\cba$).
Currently, the measurements of the Higgs boson signal strengths are compatible with the SM prediction (within the experimental uncertainties) and therefore, the parameter $\cba$ is constrained to be not far away from the alignment limit (see, for instance,~\cite{Arco:2020ucn,Arco:2022xum}).
This motivates our posterior study being classified into three qualitative different scenarios defined as:
1) alignment,  defined by setting ${\cba=0}$,  2) misalignment,  defined by arbitrary ${\cba \neq 0}$,  and 3) quasi-alignment, defined by ${\cba \ll 1 }$.  As we will see in \secref{section-matching}, the solutions for the matching will differ in these three situations. 

It is convenient to have in mind that taking the alignment condition does not imply the absence of BSM interactions, because several couplings, in addition to the SM ones,  remain non-vanishing when  $\cba=0$,  indeed,  some of them involving the light Higgs boson $h$.  For example,  the couplings $hHH$,  $hAA$,  $hH^+H^-$, $ZHA$,  $\gamma H^+H^-$,  $hhHH$,  $hhAA$, 
$hhH^+H^-$,  and others do not vanish in the alignment limit.  This implies that the integration out of the heavy modes,  $H$, $A$ and $H^\pm$ could leave some non-decoupling effects that differentiate the 2HDM with respect the SM via these couplings that could leave an important track at low energy observables,  both at tree and one-loop levels.  This is our main motivation in this work and will be discussed in the following sections.    

One interesting phenomenological feature of the 2HDM compared to the SM is the existence of triple and quartic interactions between the new scalar states.  
The 2HDM prediction of these couplings can be written in terms of the input parameters of \eqref{eq-inputs}.
Under our notation, the tree-level Feynman rules of the scalar interactions involving the light Higgs that are relevant to this work are given in \tabref{relevant-FR} of \appref{App-Frules},  and  are summarized also here:
\begin{eqnarray}
	i \Gamma_{hhh} &=& - 6 i \vev \lahhh, \label{eq-hhh} \\
	i \Gamma_{hhH} &=& - 2 i \vev \lahhH,\\
	i \Gamma_{hH^+H^-} &=& -i \vev \lahHpHm, \\
	i \Gamma_{hhhh} &=& - 6 i \lahhhh, \label{eq-hhhh}
\end{eqnarray}
where we are following the notation from~\cite{Arco:2020ucn,Arco:2022xum}.  Other scalar interactions involving the light Higgs like 
$\Gamma_{hHH}$,  $\Gamma_{hHH}$,  $\Gamma_{hAA}$,  do not participate in the present computation, and are not given explicitly here,  for shortness. 
One important aspect in this work is that these triple and quartic couplings above are not input parameters but instead they are derived parameters.  Thus,  once  the input parameters have been fixed to those in \eqref{eq-inputs},  the derived  couplings for the physical eigenstates, i.e.  the $\lambda_{\rm x}$ couplings,  are fixed in terms of the input parameters at a given order in the loop expansion.  In particular,  at the tree level,  these derived $\lambda_{\rm x}$ couplings are given by:
\begin{eqnarray}
 \label{def-lahhh}
	\vev^2\lambda_{hhh} &=& \sba\left(1+2\cba^2\right) \frac{\mh^2}{2} -\sba\cba^2\mbarc +\cba^3 \cot 2\beta \left(\mh^2-\mbarc\right) , \\
 \label{def-lahhH}
	\vev^2\lambda_{hhH} &=& \frac{\cba}{2} \left( -2 \left(3 \cba^2-2\right) \mbarc-2 \cba \sba \cot 2\beta \left(-3 \mbarc+2 \mh^2+\mHeavy^2\right) \right. \nn  \\
	&&\biggl. +\left(2\cba^2-1\right) \left(2 \mh^2+\mHeavy^2\right) \biggr), \\
 \label{def-lahHpHm}
	\vev^2\lahHpHm &=& \left(\mh^2 +2 \mHp^2 -2 \mbarc \right)\sba + 2\cot2\beta \left( \mh^2 - \mbarc \right)\cba \,,  \\
 \label{def-lahhhh} 
	\vev^2\lahhhh &=& \frac{\mh^2}{2} +\frac{\cba^2}{2}\left(-4\sba^2\mbarc +4 \cba^2 \cot^2 2\beta \left(-\mbarc+\cba^2\mh^2+\sba^2\mHeavy^2\right) \right.  \nn\\
	&&\left.+4 \cba\sba \cot 2\beta \left(-2 \mbarc+\left(2\cba^2+1\right) \mh^2+\left(1-2 \cba^2\right) \mHeavy^2\right) \right.  \nn\\
	&&\biggl.+4 \cba^4 \left(\mHeavy^2 - \mh^2\right) -4\cba^2\mHeavy^2 +3\mh^2 +\mHeavy^2\biggr).
\end{eqnarray}
where,  to present  more compact formulas we have included some derived parameters in these formulas like $\sba$, $\sb$, $\cb$ and $\cot2\beta$, that are related with the input parameters, $\cba$ and $\tan \beta$,   by the following trigonometric identities:
\be
\sba=\sqrt{1-\cba^2}\,,\,\,
\sb=\frac{\tanb}{\sqrt{1+\tanb^2}}\,, \,\,
\cb=\frac{1}{\sqrt{1+\tanb^2}}\,,\,\,
\cot2\beta=\frac{1-\tanb^2}{2\tanb}\,.
\label{trigo}
\ee
For the discussion in the following sections it is interesting to display the specific values of the couplings above in the simplest scenario with alignment,  i.e.  for ${\cba=0}$,  which are named 
$\lambda_{\rm x}^{\rm al}$ here:
\begin{eqnarray}
	\vev^2\lambda_{hhh}^{\rm al} &=&  \frac{\mh^2}{2}\,, \nn \\
		\vev^2\lambda_{hhH}^{\rm al} &=&  0\,, \nn  \\
	\vev^2\lahHpHm^{\rm al}&=& \mh^2 +2 \mHp^2 -2 \mbarc\,, \nn   \\	
	\vev^2\lahhhh^{\rm al} &=& \frac{\mh^2}{2} \,.  
\end{eqnarray}
We see  that in the alignment limit, $\lahhh$ and $\lahhhh$ tend to $\lambda=\mh^2/\left(2v^2\right)$,  which give respectively the triple and quartic Higgs couplings predicted by the SM.  The triple coupling of the two light Higgs bosons to one heavy Higgs boson vanishes in the alignment limit.  However, the coupling of one light Higgs boson to two charged Higgs bosons is not vanishing.   Regarding the above value of $\lahHpHm^{\rm al}$,  we confirm that the size of this triple coupling can be very large for large input values of $\mHp$.  For instance,  values of $\mHp \sim 800\,  {\rm GeV} $ can provide large couplings of $\lahHpHm^{\rm al}\sim {\cal O}(10)$ which, according to the  detailed analysis in~\cite{Arco:2020ucn,Arco:2022xum},  are yet allowed by all the present theoretical and experimental constraints.  This is the situation we are interested in this work.  Then,  when doing a large mass expansion of the amplitudes in the following sections we simply mean an expansion in powers of a small dimensionless parameter $\sim (v/m_{\rm heavy})$ which should be convergent whenever  $v/m_{\rm heavy} \ll 1$.  Thus,  we have in mind that our forthcoming results for heavy boson masses,   $m_H$,  $m_A$ and $m_{H^\pm}$ which are collectively named $m_{\rm heavy}$,  should apply for these masses being above $v$ and close to the TeV scale.    

Within the 2HDM, the Higgs couplings to fermions also differ with respect to the SM. 
However, since in this work we only consider interactions in the bosonic sector of the 2HDM, they are not relevant to this work and we will not describe them here.

\section{Analytical results of the amplitudes}
\label{section-amplitudes}

In this section, we present the amplitudes for the different observables considered in this work.  We focus on the following scattering and decay processes that involve the light Higgs boson and  EW gauge bosons in the external legs: 
$h\to WW^*\to Wf\bar{f'}$,  $h\to ZZ^*\to Zf\bar{f}$,  $W^+W^-\to hh$,  $ZZ\to hh$,  $hh\to hh$,  $h \to \gamma \gamma$ and $h \to \gamma Z$.  All these amplitudes were computed for the three models under consideration,  SM,  HEFT and  2HDM,  using an arbitrary $R_\xi$ gauge and verifying  the gauge parameter $\xi$ independence for the on-shell amplitudes. Thus, all the results presented here for the amplitudes are gauge invariant,  as expected. 

The following expressions were obtained using FeynArts~\cite{FeynArts} and FormCalc~\cite{FormCalc-LT}.
The relevant Feynman rules are collected in \tabref{relevant-FR} of \appref{App-Frules} (for a full set see [19–21]).

\subsection{$h\to WW^*\to Wf\bar{f'}$ and $h\to ZZ^*\to Zf\bar{f}$}

\begin{figure}[h!]
\begin{center}
\includegraphics[width=0.3\textwidth]{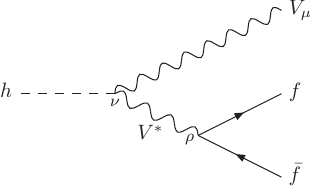}
\caption{Generic Feynman diagram  for the Higgs decays $h\to VV^*\to Vf\bar{f}$, with $V=W,Z$}
\label{fig-htoVV}
\end{center}
\end{figure}

For these decay amplitudes we follow a similar presentation as in~\cite{Anisha:2022ctm}, where these decays were also studied within the HEFT context.  We focus in this work on the tree-level amplitudes for these decays.  
We represent collectively these decay processes as $h\to VV^*\to Vf\bar{f}$,  where the EW gauge boson $V$ can be $W$ or  $Z$,  and the corresponding Feynman diagram is shown in \figref{fig-htoVV}.  The  tree-level amplitude can be generically written as:
\be
\amp = \amp^\mu\epsilon_\mu^* =(i \Gamma^{\mu\nu}_{hVV})\Delta^{VV}_{\nu\rho}(i\Gamma_{ffV}^\rho)\epsilon_\mu^*\,,
\ee
where $\epsilon_\mu^*$ is the polarization vector of the outgoing on-shell gauge boson, $\Gamma^{\mu\nu}_{hVV}$ is the 1PI with three legs corresponding to $hVV$, $\Delta^{VV}_{\nu\rho}$ is the gauge boson propagator and $\Gamma_{ffV}^\rho$ is the 1PI with three legs corresponding to $ffV$.

First,  it should be noticed that the gauge boson propagator is the same in the three considered models SM,  HEFT and 2HDM.  On the other hand,  it should also be noticed that the fermion interactions with the gauge bosons are also the same as in the SM.  
Therefore the above decay amplitudes only differ in each model on the interaction $hVV$,  which can be read from \tabref{relevant-FR}:
\bear
i \Gamma^{\mu\nu}_{hVV}\vert_{\rm SM} &=& \frac{2im_V^2}{\vev}g^{\mu\nu} \,, \nn\\
i \Gamma^{\mu\nu}_{hVV}\vert_{\rm HEFT} &=& \frac{2im_V^2}{\vev}a g^{\mu\nu} \,, \nn\\
i \Gamma^{\mu\nu}_{hVV}\vert_{\rm 2HDM} &=& \frac{2im_V^2}{\vev}\sba g^{\mu\nu}\,.
\label{vertex-hVV}
\eear
Then,  it is clear from the above expressions that the SM is recovered from the HEFT when $a=1$ (i.e.  $\Delta a =0$) and from the 2HDM in the alignment limit, i.e.  for $\cba=0$,  as expected.   Notice,  that at the tree level,  there is no dependence on the heavy Higgs boson masses in these decays.

\subsection{$W^+W^-\to hh$}
Next, we study this $WW$ scattering process at the tree level in the three considered models,  SM, HEFT and 2HDM. 
First,  to fix the notation,  we set the momenta and Lorentz indices involved in this scattering  as follows:
\be
W^+_\mu(p_+)\,W^-_\nu(p_-) \to h(k_1)\,h(k_2)\,,
\label{ourscattering-WW}
\ee
where $p_\pm$ and $k_{1,2}$ (with $p_+ +p_- =k_1 +k_2$) are the incoming and outgoing momenta of the bosons. The $W^{\pm}$ polarization vectors are $\epsilon_\pm$,  respectively. 
We present the results by separating the contributions from the various scattering channels,  $s$,  $t$,  $u$ and contact $c$ channels since we will compare the different Lorentz structures and energy dependence in our posterior study of the matching equations: 
\be
\amp = \amp\vert_s +\amp\vert_t +\amp\vert_u +\amp\vert_c \,.
\label{ampbychannels}
\ee

\begin{figure}[h!]
\begin{center}
\includegraphics[width=0.8\textwidth]{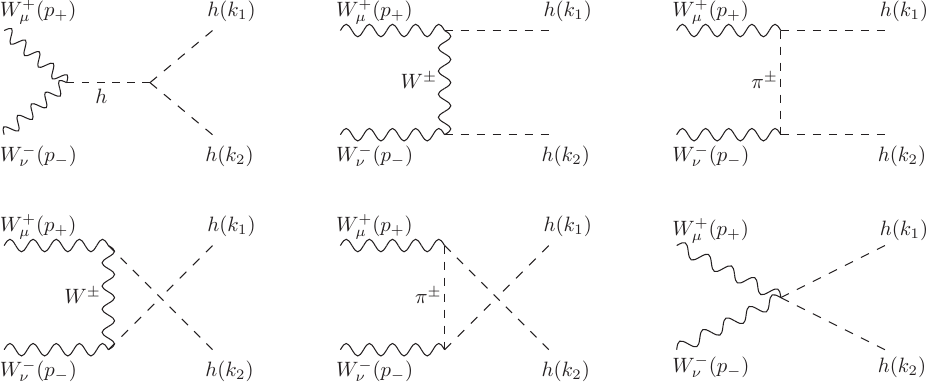}
\caption{Tree-level diagrams contributing to $WW\to hh$ in the SM and the HEFT for arbitrary $R_\xi$ gauge.}
\label{fig-WWtoHH-SMtree}
\end{center}
\end{figure}
\begin{figure}[h!]
\begin{center}
\includegraphics[width=0.8\textwidth]{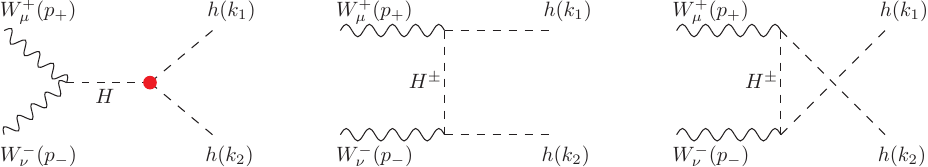}
\caption{Additional tree-level diagrams contributing to $W^+W^-\to hh$ in the 2HDM for arbitrary $R_\xi$ gauge.  The triple scalar interaction vertices of the light Higgs with heavy Higgs bosons are denoted with a big dot colored in red.}
\label{fig-WWtoHH-2HDMtree}
\end{center}
\end{figure}

Notice that the cancellation of the $\xi$ dependent terms occurs when the external  bosons are taken on shell,  and it happens  for the $t$ and $u$ channels separately.  Furthermore,  this cancellation proceeds when adding the two kind of diagrams,  with EW gauge boson and GB internal propagators that are present in both the $t$ and $u$ channels.  One can also check that the final result for the amplitude of the $R_\xi$ gauges is the same as the result obtained using the unitary gauge (with the $W$ propagator given in the unitary gauge and where no diagrams with GB modes appear in the computation),  as expected,  since the result of the amplitude must be gauge invariant.

The SM diagrams are shown in \figref{fig-WWtoHH-SMtree} and the resulting amplitude by channels is given by:
\bear
 \amp^{\rm SM} \vert_s &=& 3g^2 \frac{\lsm\vev^2}{s-\mh^2}\epsilon_{+}\cdot\epsilon_{-}\,,  \nn\\
 \amp^{\rm SM} \vert_t &=& g^2 \frac{\mw^2\epsilon_{+}\cdot\epsilon_{-} +\epsilon_{+}\cdot k_1\,\epsilon_{-}\cdot k_2}{t-\mw^2}\,,  \nn\\
 \amp^{\rm SM} \vert_u &=& g^2 \frac{\mw^2\epsilon_{+}\cdot\epsilon_{-} +\epsilon_{+}\cdot k_2\,\epsilon_{-}\cdot k_1}{u-\mw^2}\,,  \nn\\
 \amp^{\rm SM} \vert_c &=& \frac{g^2}{2} \,\epsilon_{+}\cdot\epsilon_{-}  
\label{ampWWtohh-SM}
\eear
where  $\lambda=m_h^2/(2v^2)$.

Within the HEFT the diagrams are also collected in \figref{fig-WWtoHH-SMtree} and the result for the amplitude is also gauge invariant.  The corresponding LO contributions (i.e.  from from $\mL_2$) from the various channels, at the tree level,  are given by:
\bear
\amp^{\rm HEFT}\vert_s &=& 3g^2a\kappa_3\frac{\lsm\vev^2}{s-\mh^2}\epsilon_{+}\cdot\epsilon_{-} \,, \nn\\
\amp^{\rm HEFT}\vert_t &=& g^2a^2\frac{\mw^2\epsilon_{+}\cdot\epsilon_{-} +\epsilon_{+}\cdot k_1\,\epsilon_{-}\cdot k_2}{t-\mw^2} \,, \nn\\
\amp^{\rm HEFT}\vert_u &=& g^2a^2\frac{\mw^2\epsilon_{+}\cdot\epsilon_{-} +\epsilon_{+}\cdot k_2\,\epsilon_{-}\cdot k_1}{u-\mw^2}\,,  \nn\\
\amp^{\rm HEFT}\vert_c &=& \frac{g^2}{2}b\,\epsilon_{+}\cdot\epsilon_{-} \,.
\label{ampWWtohh-EChL}
\eear
where again the relation $\lambda=\mh^2/(2\vev^2)$ is understood.

Regarding the prediction of the amplitude  for the 2HDM in covariant $R_\xi$  gauges,   we notice that in addition to the SM-like diagrams in  \figref{fig-WWtoHH-SMtree},  where the interchanged Higgs boson is the light Higgs $h$,  there are also the contributions from the exchange of a heavy neutral Higgs boson in the $s$-channel and from the heavy charged Higgs bosons  in the $t$- and $u$-channels, as it is shown in \figref{fig-WWtoHH-2HDMtree}. 
Notice also that the contributions from all these diagrams are $\xi$ independent.  And the total result for the amplitude is again gauge invariant.  
The corresponding amplitudes by channels are:
\bear
 \amp^{\rm 2HDM} \vert_s &=& g^2 \left(\frac{3\lambda_{hhh}\vev^2}{s-\mh^2}\sba+\frac{\lambda_{hhH}\vev^2}{s-\mHeavy^2}\cba\right) \epsilon_{+}\cdot\epsilon_{-} \,, \nn\\
 \amp^{\rm 2HDM} \vert_t &=& g^2 \sba^2 \frac{\mw^2\epsilon_{+}\cdot\epsilon_{-} +\epsilon_{+}\cdot k_1\,\epsilon_{-}\cdot k_2}{t-\mw^2} +g^2\cba^2\frac{\epsilon_{+}\cdot k_1\,\epsilon_{-}\cdot k_2}{t-\mHp^2} \,, \nn\\
 \amp^{\rm 2HDM} \vert_u&=& g^2 \sba^2 \frac{\mw^2\epsilon_{+}\cdot\epsilon_{-} +\epsilon_{+}\cdot k_2\,\epsilon_{-}\cdot k_1}{u-\mw^2} +g^2\cba^2\frac{\epsilon_{+}\cdot k_2\,\epsilon_{-}\cdot k_1}{u-\mHp^2} \,, \nn\\
 \amp^{\rm 2HDM} \vert_c &=& \frac{g^2}{2} \,\epsilon_{+}\cdot\epsilon_{-}  \,.
\label{ampWWtohh-2HDM}
\eear
In the $s$ channel, we have used the compact notation for the derived triple Higgs couplings,   $\lambda_{hhh}$ and 
$\lambda_{hhH}$ of \eqrefs{def-lahhh}{def-lahhH}.

It is important to remark that the SM predictions of \eqref{ampWWtohh-SM} are recovered from the HEFT ones of \eqref{ampWWtohh-EChL} by taking $a=b=\kappa_3=1$ (i.e.  for $\Delta a=\Delta b = \Delta \kappa_3=0$).  On the other hand,  we have also checked that the 2HDM results of \eqref{ampWWtohh-2HDM} in the alignment limit (i.e.  for $\cba=0$) coincide with the SM ones. 
In \secref{section-matching} we will analyze the effect of the heavy Higgs boson masses  away from the alignment limit.

\subsection{$ZZ\to hh$}
\begin{figure}[h!]
\begin{center}
\includegraphics[width=0.8\textwidth]{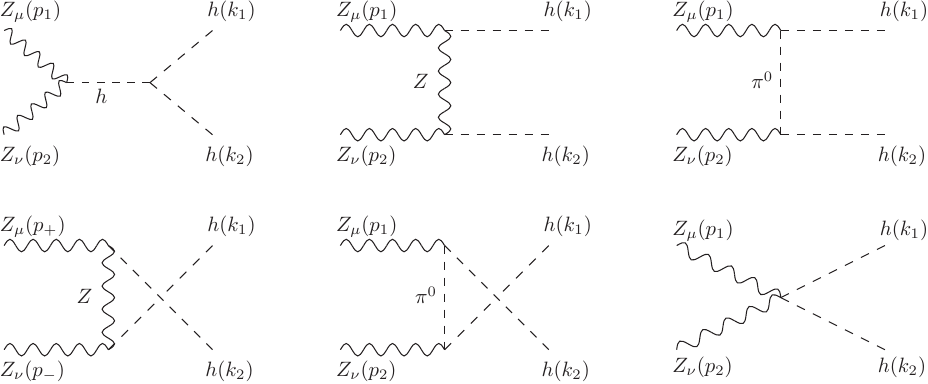}
\caption{Tree-level diagrams contributing to $ZZ\to hh$ in the SM and the HEFT for arbitrary $R_\xi$ gauge.}
\label{fig-ZZtoHH-SMtree}
\end{center}
\end{figure}

\begin{figure}[h!]
\begin{center}
\includegraphics[width=0.8\textwidth]{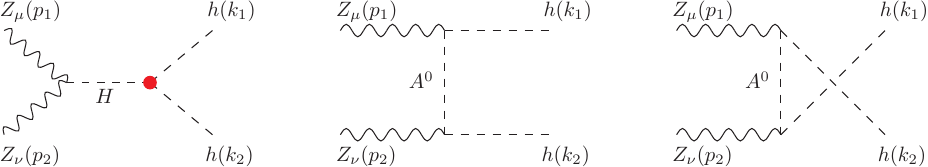}
\caption{Additional tree-level diagrams contributing to $ZZ\to hh$ in the 2HDM for arbitrary $R_\xi$ gauge.  The triple scalar interaction vertices of the light Higgs with heavy Higgs bosons are denoted with a big dot colored in red.}
\label{fig-ZZtoHH-2HDMtree}
\end{center}
\end{figure}

We follow a similar presentation here as in the previous process. 
The corresponding notation for the momenta in this case is:
\be
Z_\mu(p_1)\,Z_\nu(p_2) \to h(k_1)\,h(k_2)\,,
\label{ourscattering-ZZ}
\ee
where $p_{1,2}$ and $k_{1,2}$ (with $p_1 +p_2 =k_1 +k_2$) are the incoming and outgoing momenta of the bosons. The $Z$ polarization vectors are $\epsilon_{1,2}$,  respectively.
The resulting diagrams in the SM and EChL are collected in \figref{fig-ZZtoHH-SMtree} and the additional diagrams in the 2HDM are shown in \figref{fig-ZZtoHH-2HDMtree}.

Notice  that the amplitudes in this case can also be derived from the previous ones by multiplying them by $\cw^{-2}$ and replacing $\mw\to\mz$ and $\mHp\to\mA$.  Explicitly,  the SM amplitude by channels is:
\bear
 \amp^{\rm SM} \vert_s &=& 3\frac{g^2}{\cw^2} \frac{\lsm\vev^2}{s-\mh^2}\epsilon_{1}\cdot\epsilon_{2} \,, \nn\\
 \amp^{\rm SM} \vert_t&=& \frac{g^2}{\cw^2} \frac{\mz^2\epsilon_{1}\cdot\epsilon_{2} +\epsilon_{1}\cdot k_1\,\epsilon_{2}\cdot k_2}{t-\mz^2} \,, \nn\\
 \amp^{\rm SM} \vert_u &=& \frac{g^2}{\cw^2} \frac{\mz^2\epsilon_{1}\cdot\epsilon_{2} +\epsilon_{1}\cdot k_2\,\epsilon_{2}\cdot k_1}{u-\mz^2} \,, \nn\\
 \amp^{\rm SM} \vert_c &=& \frac{g^2}{2\cw^2} \,\epsilon_{1}\cdot\epsilon_{2}  \,.
\label{ampZZtohh-SM}
\eear

The HEFT amplitude by channels is:
\bear
\amp^{\rm HEFT}\vert_s&=& 3\frac{g^2}{\cw^2}a\kappa_3\frac{\lsm\vev^2}{s-\mh^2}\epsilon_{1}\cdot\epsilon_{2} \,,  \nn\\
\amp^{\rm HEFT}\vert_t &=& \frac{g^2}{\cw^2}a^2\frac{\mz^2\epsilon_{1}\cdot\epsilon_{2} +\epsilon_{1}\cdot k_1\,\epsilon_{2}\cdot k_2}{t-\mz^2} \,, \nn\\
\amp^{\rm HEFT}\vert_u&=& \frac{g^2}{\cw^2}a^2\frac{\mz^2\epsilon_{1}\cdot\epsilon_{2} +\epsilon_{1}\cdot k_2\,\epsilon_{2}\cdot k_1}{u-\mz^2} \,, \nn\\
\amp^{\rm HEFT}\vert_c &=& \frac{g^2}{2\cw^2}b\,\epsilon_{1}\cdot\epsilon_{2} \,.
\label{ampZZtohh-EChL}
\eear

The 2HDM amplitude by channels is:
\bear
 \amp^{\rm 2HDM} \vert_s &=& \frac{g^2}{\cw^2} \left(\frac{3\vev^2}{s-\mh^2}\lambda_{hhh}\sba+\frac{\vev^2}{s-\mHeavy^2}\lambda_{hhH}\cba\right) \epsilon_{1}\cdot\epsilon_{2} \,, \nn\\
 \amp^{\rm 2HDM} \vert_t &=& \frac{g^2}{\cw^2} \sba^2 \frac{\mz^2\epsilon_{1}\cdot\epsilon_{2} +\epsilon_{1}\cdot k_1\,\epsilon_{2}\cdot k_2}{t-\mz^2} +\frac{g^2}{\cw^2}\cba^2\frac{\epsilon_{1}\cdot k_1\,\epsilon_{2}\cdot k_2}{t-\mA^2} \,, \nn\\
 \amp^{\rm 2HDM} \vert_u&=& \frac{g^2}{\cw^2} \sba^2 \frac{\mz^2\epsilon_{1}\cdot\epsilon_{2} +\epsilon_{1}\cdot k_2\,\epsilon_{2}\cdot k_1}{u-\mz^2} +\frac{g^2}{\cw^2}\cba^2\frac{\epsilon_{1}\cdot k_2\,\epsilon_{2}\cdot k_1}{u-\mA^2} \,, \nn\\
 \amp^{\rm 2HDM} \vert_c &=& \frac{g^2}{2\cw^2} \,\epsilon_{1}\cdot\epsilon_{2}  \,.
\label{ampZZtohh-2HDM}
\eear

In particular, the SM results are also recovered from the HEFT ones when $a=b=\kappa_3=1$  (i.e.  for $\Delta a=\Delta b = \Delta \kappa_3=0$) and from the 2HDM in the alignment limit (i.e.  for $\cba=0$).

\subsection{$hh\to hh$}

\begin{figure}[h!]
\begin{center}
\includegraphics[width=0.6\textwidth]{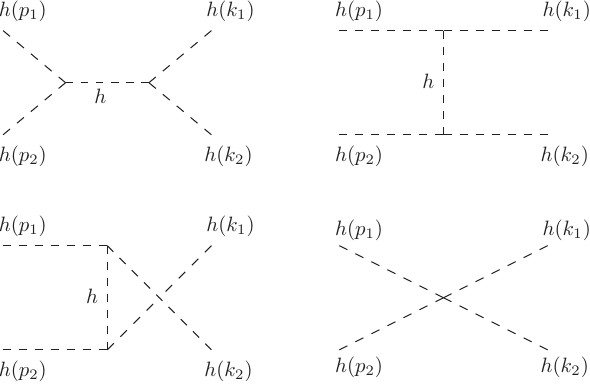}
\caption{Tree-level diagrams contributing to $hh\to hh$ in the SM and in the HEFT for arbitrary $R_\xi$ gauge.}
\label{fig-hhtohh-SMtree}
\end{center}
\end{figure}

\begin{figure}[h!]
\begin{center}
\includegraphics[width=0.8\textwidth]{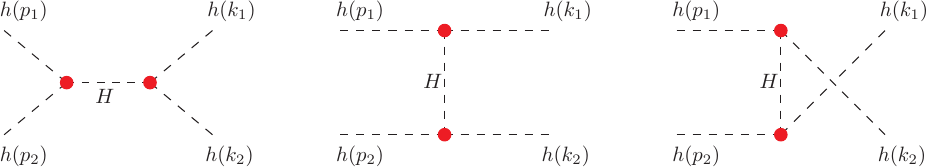}
\caption{Additional tree-level diagrams contributing to $hh\to hh$ in the 2HDM for arbitrary $R_\xi$ gauge.  The triple scalar interaction vertices of the light Higgs with heavy Higgs bosons are denoted with a big dot colored in red.}
\label{fig-hhtohh-2HDMtree}
\end{center}
\end{figure}

The tree-level diagrams contributing in the SM and the HEFT are the same set and are collected in \figref{fig-hhtohh-SMtree}. In the case of the 2HDM,  these four diagrams in  \figref{fig-hhtohh-SMtree} also contribute,  and besides them there are also diagrams where the heavy neutral Higgs boson propagates in the $s$-, $t$- and $u$-channels.  These additional diagrams are collected  in \figref{fig-hhtohh-2HDMtree}.
It should be noted that all these diagrams are independent of the $\xi$ gauge parameter, and therefore the result for the amplitudes in the three models are gauge invariant, as expected. 

The SM amplitude by channels is:
\bear
 \amp^{\rm SM} \vert_s &=& -\frac{36\lsm^2\vev^2}{s-\mh^2} \,, \nn\\
 \amp^{\rm SM} \vert_t &=& -\frac{36\lsm^2\vev^2}{t-\mh^2} \,, \nn\\
 \amp^{\rm SM} \vert_u &=& -\frac{36\lsm^2\vev^2}{u-\mh^2} \,, \nn\\
 \amp^{\rm SM} \vert_c &=& -6\lsm \,,
\label{amphhtohh-SM}
\eear
where,  again,  $ \lambda=m_h^2/ (2 v^2)$.  

The HEFT amplitude by channels is:
\bear
 \amp^{\rm HEFT} \vert_s &=& -\frac{36\lsm^2\vev^2}{s-\mh^2}\kappa_3^2 \,, \nn\\
 \amp^{\rm HEFT} \vert_t &=& -\frac{36\lsm^2\vev^2}{t-\mh^2}\kappa_3^2 \,, \nn\\
 \amp^{\rm HEFT} \vert_u &=& -\frac{36\lsm^2\vev^2}{u-\mh^2}\kappa_3^2 \,, \nn\\
 \amp^{\rm HEFT} \vert_c &=& -6\lsm\kappa_4 \,.
\label{amphhtohh-EChL}
\eear

The 2HDM amplitude by channels is:
\bear
 \amp^{\rm 2HDM} \vert_s &=& -\frac{36\vev^2}{s-\mh^2}\lahhh^2 -\frac{4\vev^2}{s-\mHeavy^2}\lahhH^2 \,, \nn\\
 \amp^{\rm 2HDM} \vert_t &=& -\frac{36\vev^2}{t-\mh^2}\lahhh^2 -\frac{4\vev^2}{t-\mHeavy^2}\lahhH^2 \,, \nn\\
 \amp^{\rm 2HDM} \vert_u &=& -\frac{36\vev^2}{u-\mh^2}\lahhh^2 -\frac{4\vev^2}{u-\mHeavy^2}\lahhH^2 \,, \nn\\
 \amp^{\rm 2HDM} \vert_c &=& -6\lahhhh \,,
\label{amphhtohh-2HDM}
\eear
where the derived values for the triple $\lahhh$,  $\lahhH$  and quartic $\lahhhh$  couplings are given in 
 \eqrefs{def-lahhh}{def-lahhH} and \eqref{def-lahhhh},  respectively.

As in the previous observables, the SM results are recovered from the HEFT ones for $\kappa_3=\kappa_4=1$ (i.e.  for $ \Delta \kappa_3=  \Delta \kappa_4=0$) and from the 2HDM in the alignment limit (i.e.  for $\cba=0$). 

\subsection{$h\to\gamma\gamma$}
This decay occurs at one-loop level in the three considered models,  thus the predictions for their corresponding  amplitudes are all of ${\cal O}(\hbar/(16 \pi^2))$.   
The computation of this observable in $R_\xi$ covariant gauges was presented in~\cite{Marciano:2011gm} for the SM and in~\cite{paperHdecays} for the HEFT,  where it was also recomputed the SM case,  for comparison.  In this presentation we follow the computation as described in~\cite{paperHdecays}.  

\begin{figure}[h!]
\begin{center}
\includegraphics[width=0.8\textwidth]{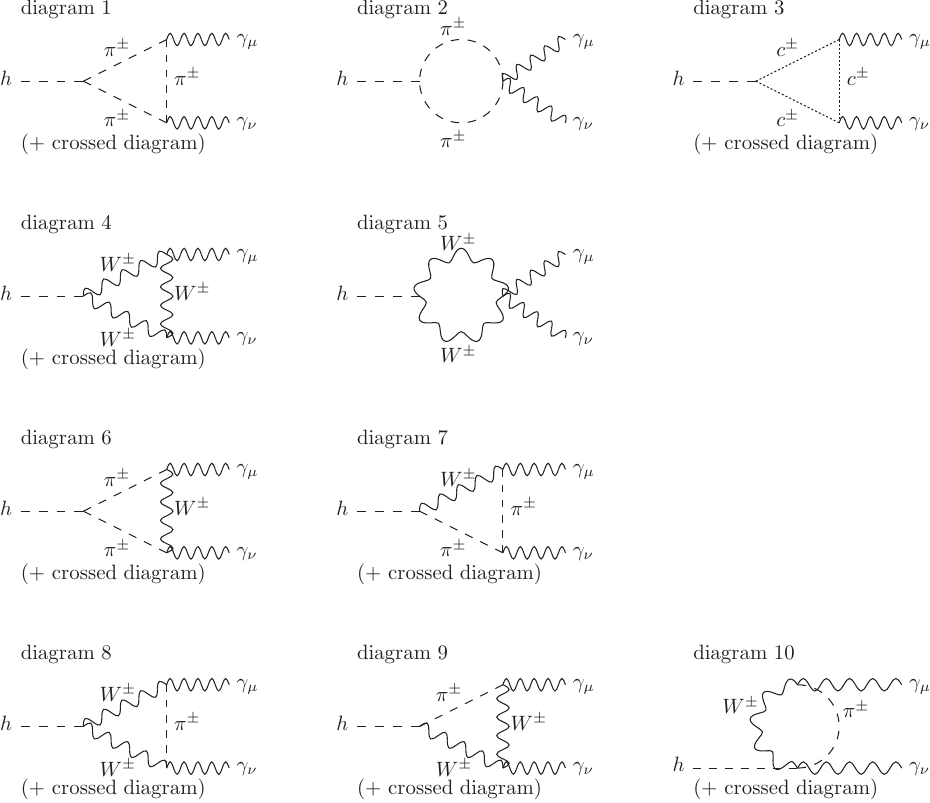}
\caption{Bosonic one-loop diagrams contributing to $h\to\gamma\gamma$ within the SM in covariant $R_\xi$ gauges.  
 In the HEFT case,  these one-loop diagrams also contribute to the  amplitude.  The contributions from these diagrams are different in the SM and HEFT cases.   In the 2HDM case,   these one-loop diagrams also contribute,  but again with different values than in the SM and in the HEFT.  Within the HEFT there are additional diagrams as summarized in \figref{fig-HEFT-htogaga}.  Within the 2HDM there are additional diagrams as summarized in \figref{fig-2HDM-htogaga}.}
\label{fig-SM-htogaga}
\end{center}
\end{figure}

In the SM case,  when adding all the 1-loop diagrams in the $R_\xi$  gauges given in \figref{fig-SM-htogaga},  the total one-loop amplitude is UV finite and does not need for renormalization nor counterterms.  The $\xi$-dependence cancellation among the various loop diagrams, leading to the gauge invariance of the resulting one-loop amplitude was also fully discussed in~\cite{paperHdecays}.  Notice that for the purpose of the present work where we are interested only in the bosonic part of the models,  we do not need to compute the loops with fermions.  

The result of the SM one-loop amplitude for the decay $h(q)\to\gamma(k_1)\gamma(k_2)$ in covariant $R_\xi$ gauges respects the Lorentz structure expected by the $U(1)$ Ward identity and can be written as:
\be
\amp^{\rm SM}(h\to\gamma\gamma)=\frac{1}{\vev}F_{h\gamma\gamma} \left(\mh^2 (\epsilon_1 \epsilon_2)  -2 (\epsilon_1 k_2)(\epsilon_2 k_1) \right) \,,
\ee 
where the explicit computation of all the bosonic loops gives:
\be
F_{h\gamma\gamma}= \frac{g^2 \sw^2}{8 \pi ^2 \mh^2} \left(12 \frac{\mw^2}{\mh^2} \left(\mh^2-2 \mw^2\right)f\left(\xh\right)+\mh^2+6 \mw^2\right) \,.
\ee
The definition of the \1loop function $f(r)$ in the above formula is given in \appref{App-floops}.  The result has also been checked to be the same when computing the loop contributions to the amplitude  in the unitary gauge (see also,~\cite{paperHdecays}).  

\begin{figure}[h!]
\begin{center}
\includegraphics[width=0.8\textwidth]{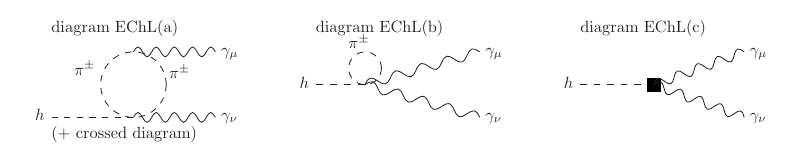}
\caption{Additional diagrams contributing to $h\to\gamma\gamma$ in the HEFT for arbitrary $R_\xi$ gauge. The black box represents the contribution from $a_{h\gamma\gamma}$ in $\mL_4$. }
\label{fig-HEFT-htogaga}
\end{center}
\end{figure}

For the HEFT case,  the same one-loop diagrams displayed in \figref{fig-SM-htogaga} contribute to the decay amplitude but with different values than in the SM.  Recall that the Feynman rules in  both theories are different.  In particular,  the ghost diagrams are absent within the HEFT (i.e.  diagram 3 of this figure vanishes) due to the vanishing couplings of the $h$ with the ghosts in this non-linear theory (see~\cite{paperHdecays} for details and also for the interesting comparison between the computation in the covariant gauges and the unitary gauge).  The additional diagrams in the HEFT are represented in \figref{fig-HEFT-htogaga}.   All the loop diagrams come from $\mL_2$ of \eqref{eq-L2} and the tree-level contribution  comes from the $\mL_4$ of \eqref{eq-L4relevant}.  Diagrams (a) and (b) appear in the HEFT due to the non-linear representation for the GBs that provide new couplings for them that are not present in the SM.  Diagram (c) appears due to the tree-level contributions from the effective operators in $\mL_4$ which provide a term in the decay amplitude involving the effective coefficient  $a_{h\gamma\gamma}$.   It is interesting to recall that  the sum all the one-loop diagrams in the HEFT gives also a UV finite result. This means that $a_{h\gamma\gamma}$
does not need to be renormalized,  i.e. in this $h$ decay the relevant coefficient in $\mL_4$ does not need to act as a counterterm for the loops from $\mL_2$ since they provide a finite result. 

The result of the HEFT amplitude for this decay is:
\be
\amp^{\rm HEFT}(h\to\gamma\gamma)=\amp^{\mL_2^{\rm loop}}+\amp^{\mL_4^{\rm tree}}
= \frac{1}{\vev}\left( aF_{h\gamma\gamma}+g^2\sw^2a_{h\gamma\gamma} \right)
 \left(\mh^2 (\epsilon_1 \epsilon_2)  -2 (\epsilon_1 k_2)(\epsilon_2 k_1) \right)\,,
\ee
where $a\,F_{h\gamma\gamma}$ collects the contributions  from the loops computed with $\mL_2$ and the term proportional to $a_{h\gamma\gamma}$ corresponds to the tree-level contribution from $\mL_4$.

\begin{figure}[h!]
\begin{center}
\includegraphics[width=0.6\textwidth]{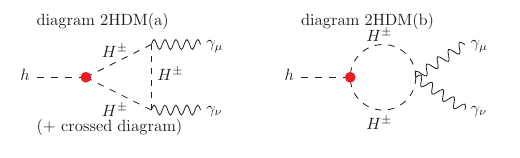}
\caption{Additional  diagrams contributing to $h\to\gamma\gamma$ in the 2HDM for arbitrary $R_\xi$ gauge. The triple scalar interaction vertices of the light Higgs with heavy Higgs bosons are denoted with a big dot colored in red.}
\label{fig-2HDM-htogaga}
\end{center}
\end{figure} 

The computation of the one-loop amplitude in  the 2HDM  case for the generic $R_\xi$ gauges,   involves the computation~\cite{HHG,Fontes:2014xva} of the  one-loop diagrams in \figref{fig-SM-htogaga},  and the additional one-loop diagrams in \figref{fig-2HDM-htogaga}.  We have checked that the sum of all loops are $\xi$ independent. Notice that these additional diagrams correspond to the one-loop contributions involving the charged Higgs bosons in the internal propagators and  are $\xi$-independent.   We have checked that after adding all the loop diagrams the result is also  $\xi$ independent. 
The resulting amplitude (bosonic contributions) within the 2HDM is the following:
\begin{equation}
\amp^{\rm 2HDM}(h\to\gamma\gamma) = \frac{1}{\vev}\left( \sba F_{h\gamma\gamma} +F_{h\gamma\gamma}^{H^\pm} \right)
 \left(\mh^2 (\epsilon_1 \epsilon_2)  -2 (\epsilon_1 k_2)(\epsilon_2 k_1) \right),
\end{equation}
where $\sba\,F_{h\gamma\gamma}$ corresponds to the diagrams  in \figref{fig-SM-htogaga}
and $F_{h\gamma\gamma}^{H^\pm}$ is the extra contribution from the $H^\pm$ loops in \figref{fig-2HDM-htogaga}. This later is given by:
\begin{equation}
F_{h\gamma\gamma}^{H^\pm}=\frac{g^2\vev^2\lahHpHm\sw^2 }{8\pi^2\mh^2}\left(1 - \frac{4\mHp^2}{\mh^2}f\left( \frac{4\mHp^2}{\mh^2} \right)\right)\,.
\label{hAAloop-Hp}
\end{equation} 
Again the function $f(r)$ is given in \appref{App-floops}.
An important point to remark here is that $F_{h\gamma\gamma}^{H^\pm}$ depends on the triple coupling $\lahHpHm$ whose derived value in terms of the input  parameters is given in \eqref{def-lahHpHm}.

\subsection{$h\to\gamma Z$}

\begin{figure}[h!]
\begin{center}
\includegraphics[width=0.6\textwidth]{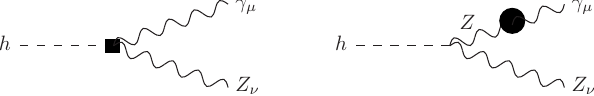}
\caption{Additional  diagrams contributing to $h\to\gamma Z$ that are not present in $h\to\gamma\gamma$. The black box represents the tree-level contribution of the counterterms. The big black ball represents the renormalized 2-point 1PI $\gamma Z$ function.}
\label{fig-htogaZ}
\end{center}
\end{figure}

The computation of the decay amplitude in the $h \to \gamma Z$  is similar to the previous $h \to \gamma \gamma$ case (with slight differences commented below) and it was  also presented in~\cite{paperHdecays} for arbitrary $R_\xi$ gauge within both SM and HEFT.  We recall here just the most relevant points. 
The main set of one-loop diagrams in the SM are the same as in \figref{fig-SM-htogaga} replacing one photon by a $Z$ boson.  In this case, however, the sum of all the one-loop diagrams in this figure is not UV convergent and the contributions  in \figref{fig-htogaZ} must be included in order to get a UV finite and $\xi$-independent amplitude respecting the Ward identity.
In particular, we include the counterterm of the 3-legs 1PI Green function (left diagram in \figref{fig-htogaZ}) and the loop and counterterm contributions yielding to the 2-legs renormalized 1PI self-energy photon-$Z$ (represented by the black ball in the right diagram of this figure).

The result of the SM one-loop amplitude (bosonic contributions) for $h(q)\to\gamma(k_1)Z(k_2)$  in a $R_\xi$ gauge  is the following: 
\be
\amp^{\rm SM}(h\to\gamma Z)=\frac{1}{\vev}F_{h\gamma Z} \left((\mh^2-\mz^2) (\epsilon_1 \epsilon_2)  -2 (\epsilon_1 k_2)(\epsilon_2 k_1) \right)\,,
\ee
where 
\bear
F_{h\gamma Z} &=& \frac{g^2\sw\cw}{16\pi^2(\mh^2-\mz^2)}  \left( 2\mh^2+12\mw^2-\frac{\mz^2}{\mw^2}\mh^2-2\mz^2 \right.\nn\\
&&\left. \quad\hspace{5mm}-\frac{4\mw^2}{\mh^2-\mz^2}\left(-6\mh^2+12\mw^2+\frac{\mz^2}{\mw^2}\mh^2+6\mz^2-\frac{2\mz^4}{\mw^2}\right) \left(f\left(\xh\right)-f\left(\xz\right)\right) \right.\nn\\
&&\left. \quad\hspace{5mm}-\frac{2\mz^2}{\mh^2-\mz^2}\left(2\mh^2+12 \mw^2-\frac{\mz^2}{\mw^2}\mh^2-2\mz^2\right)\left(g\left(\xh\right)-g\left(\xz\right)\right)\right)\,.
\eear
The definition of the \1loop functions $f(r)$ and $g(r)$ are given in \appref{App-floops}.

The result of the HEFT one-loop amplitude (bosonic contributions) for $h(q)\to\gamma(k_1)Z(k_2)$  in a $R_\xi$ gauge  is the following: 
\be
\amp^{\rm HEFT}(h\to\gamma Z)=\amp^{\mL_2^{\rm loop}}+\amp^{\mL_4^{\rm tree}}
=\frac{1}{\vev}\left( a\,F_{h\gamma Z} +g^2\sw\cw a_{h\gamma Z} \right)
 \left((\mh^2-m_Z^2) (\epsilon_1 \epsilon_2)  -2 (\epsilon_1 k_2)(\epsilon_2 k_1) \right)\,,
\ee
where $a\,F_{h\gamma Z}$ summarizes  the one-loop contributions from $\mL_2$ and the term proportional to 
$a_{h\gamma Z}$ comes from $\mL_4$ at the tree level.  In this case,   the coefficient $a_{h\gamma Z}$ also acts as counterterm. 

The result of the 2HDM one-loop amplitude~\cite{HHG,Fontes:2014xva} (bosonic contributions) for $h(q)\to\gamma(k_1)Z(k_2)$  in a $R_\xi$ gauge  is the following: : 
\begin{equation}
\amp^{\rm 2HDM}(h\to\gamma Z)=\frac{1}{\vev}\left( \sba F_{h\gamma Z} +F_{h\gamma Z}^{H^\pm} \right)
 \left((\mh^2-\mz^2) (\epsilon_1 \epsilon_2)  -2 (\epsilon_1 k_2)(\epsilon_2 k_1) \right),
\end{equation}
where $\sba F_{h\gamma Z}$ comes from the  one-loop diagrams and counterterms already present in the SM and $F_{h\gamma Z}^{H^\pm}$ is the additional contribution from the $H^\pm$ loops:
\bear
F_{h\gamma Z}^{H^\pm} &=& \frac{\lahHpHm\sw\cw(2\mw^2-\mz^2)}{4\pi^2(\mh^2-\mz^2)^2}\left(\mh^2-\mz^2 -2\mz^2\left(g\left(\frac{4\mHp^2}{\mh^2}\right)-g\left(\frac{4\mHp^2}{\mz^2}\right)\right)  \right.  \nn\\
&&\left. \hspace{4mm}-\mHp^2\log^2\left(\frac{2\mHp^2-\mz^2+\sqrt{-\mz^2(4\mHp^2-\mz^2)}}{2\mHp^2}\right) \right.  \nn\\
&&\left. \hspace{4mm}+\mHp^2\log^2\left(\frac{2\mHp^2-\mh^2+\sqrt{-\mh^2(4\mHp^2-\mh^2)}}{2\mHp^2}\right) \right) \,.
\label{hAZloop-Hp}
\eear
Once again, this contribution is proportional to the triple  coupling of the light Higgs with the charged Higgs bosons \eqref{def-lahHpHm}.

\section{Matching HEFT and 2HDM amplitudes}
\label{section-matching}
In this section,  we first set the matching equations relating  the HEFT and the 2HDM amplitudes,  and second we solve these equations analytically providing the solutions for the HEFT coefficients in terms of the 2HDM input parameters.
Firstly, we define this matching by equating the amplitudes from the HEFT with the amplitudes  from the 2HDM in the heavy Higgs bosons limit.  This heavy Higgs boson limit refers to consider heavy $H$,  $A$ and $H^{^\pm}$ with respect to the EW scale $v$ or, equivalently,  respect to all EW masses involved: $ \mHeavy,\, m_A,  \mHp \gg \mw,\,\mz,\,\mh,\,\vev,\,m_{12}$. 
In the case of scattering amplitudes,  by heavy  BSM Higgs boson limit we also mean heavy respect to the energy of the scattering process,  i.e.,  $m_{\rm heavy} \gg \sqrt{s}$,  where  $m_{\rm heavy}$ denotes collectively any of the heavy masses involved,    
$\mHeavy,\, m_A,  \mHp$.  
Thus,  for shortness , in the following,  we will refer generically to this large heavy mass limit,  by requiring a big hierarchy among the two scales, i.e., by considering:
\be
m_{\rm heavy} \gg m_{\rm EW} \,.
\ee 
Where,  generically,   we are assuming  that $m_{\rm heavy}$ is  closer to the TeV scale and  $m_{\rm EW}$,  being  the masses and energies involved,  is closer to the EW scale. 
Then,  for a given amplitude $\amp$  our generic matching condition reads as follows:
\be
\amp^{\rm HEFT}=\amp^{\rm 2HDM}_{\rm heavy} \,,
\label{eq-ampmatching}
\ee
where  $\amp^{\rm HEFT}$ is the prediction from the HEFT and $\amp^{\rm 2HDM}_{\rm heavy}$ means the result of the 2HDM amplitude after integrating out the heavy modes.  This integration is performed in practice by means of an expansion of the 2HDM amplitude in inverse powers of the heavy boson masses.  Generically, this large mass expansion will lead to terms in the total amplitude with increasing powers in a small mass ratio,  namely,  with increasing  orders in $\sim (m_{\rm EW}^2/m_{\rm heavy}^2)^n$ with $n=0,2, 4, ...$.  On the other hand,  this matching condition can be set at any order in the loop expansion,  i.e. to ${\cal O}(\hbar^0)$,  ${\cal O}(\hbar^1)$, etc,  but in any case this order must be fixed equally in both sizes of the matching equations.  Correspondingly,  the solutions to the matching equations will be provided at a fixed order, i.e.  either at the tree level,  or at the one-loop level etc.  Finally,  since we are mainly interested in capturing the non-decoupling effects from the heavy Higgs bosons,  encoded in those solutions,  we will need to focus only on the leading terms of this large mass expansion. Therefore,  we will keep in this work only 
those leading terms in  $\amp^{\rm 2HDM}_{\rm heavy}$ that go with the $n=0$ power, or equivalently,  the contributions which are constant with the heavy mass $m_{\rm heavy}$ in the heavy mass limit.   As we will see, there are no contributions with negative $n$, i.e.,  the total result of the amplitude never grows with the heavy masses,  thus demonstrating that the large mass expansion is convergent.  The final comment regarding the previous matching equation is that when solving it, we must write the solutions for BSM with respect to the SM ones. This is equivalent to say that the solutions of the matching equations at the leading order in the large mass expansion,  must be provided at the end for $\Delta a$,  $\Delta b$,  $\Delta {\kappa_3}$,  
$\Delta {\kappa_4}$,  $a_{h\gamma \gamma}$ and $a_{h \gamma Z}$,  and these must be given  in terms of the input parameters (other than the heavy masses) of the 2HDM,   namely, in terms of $\vev$, $\mh$, $\cba$, $\tanb$ and $m_{12}$,  as set in \eqref{eq-inputs}.

Our matching procedure systematically solves the above \eqref{eq-ampmatching} by taking into account all the Lorentz structures involved and considering both  the energy and scattering angle dependencies of the scattering  amplitudes,  $\amp=\amp(s,\theta)$ (or masses and momenta  dependence  in the case of decay amplitudes).  Then we solve the matching conditions sequentially and process by process. 
In this matching we consider the full set of seven amplitudes  presented in the previous section.  Notice that some HEFT coefficients are present in various amplitudes, therefore once we solve them from one subset of amplitudes, the remaining ones are used to cross-check the results.   This is useful to check that our solutions for the HEFT coefficients in terms of the 2HDM input parameters are the same for all the considered processes, and therefore they are  process-independent results.

\subsection{Amplitudes for  large $m_{\rm heavy}$ in the 2HDM and solutions to the matching}
As announced above,  we keep here just the dominant contributions in the large  $m_{\rm heavy}$ expansion that  define the non-decoupling effects from the heavy modes,  which for the selected amplitudes here correspond to the resulting contributions that are independent of the heavy Higgs boson masses, i.e. they behave as $(m_{\rm EW}/m_{\rm heavy})^0$.  The next to leading terms in these expansions are decoupling,  going as 
$(m_{\rm EW}/m_{\rm heavy})^n$ with $n=2,4...$ etc.,  and are not included in our solutions to the matching equations.  In the computation of this expansion it is crucial to 
take into account that $\lahhH$, $\lahHpHm$ and $\lahhhh$ depend on the heavy masses of the BSM Higgs bosons as it is shown in \eqrefs{def-lahhH}{def-lahhhh}.  Another important point to take into account is that,  due to the previously shown $\xi$ independence of the separate contributions to the amplitudes from the various channels $s, t, u$ and $c$,  the matching equations can be analyzed by channels. 

{\bf 1) $h\to VV^*\to Vf\bar{f}$} 

The starting matching equation is for the decay amplitudes of the light Higgs boson  into a gauge boson $W$ or $Z$ and a fermion pair,  at the tree level.  The solution of the matching equation in this case is trivial since the 2HDM amplitude does not depend on $m_{\rm heavy}$.  
Therefore, the solution to the matching equation in these two decays gives simply:
\be
a =1-\Delta a= \sba \,.
\label{matching-a}
\ee

{\bf 2) $W^+W^- \to hh$} 

The next matching equation is for the $W^+W^- \to hh$ scattering.  In this case,  the large heavy mass expansion gives the following results for the 2HDM amplitude,  presented here by channels:
\bear
 \amp^{\rm 2HDM}_{\rm heavy} \vert_s &=& g^2 \left(\frac{3\sba}{s-\mh^2}\left(\sba\left(1+2\cba^2\right) \frac{\mh^2}{2} -\sba\cba^2\mbarc +\cba^3 \cot 2\beta \left(\mh^2-\mbarc\right)\right) \right.  \nn\\
 &&\left.\hspace{2mm}-\frac{\cba^2}{2} \left( -2 \cba \sba \cot 2\beta  +2\cba^2-1 \right)\right) \epsilon_{+}\cdot\epsilon_{-} \,, \nn\\
 \amp^{\rm 2HDM}_{\rm heavy}  \vert_t &=& g^2 \sba^2 \frac{\mw^2\epsilon_{+}\cdot\epsilon_{-} +\epsilon_{+}\cdot k_1\,\epsilon_{-}\cdot k_2}{t-\mw^2} \,, \nn\\
 \amp^{\rm 2HDM}_{\rm heavy}  \vert_u &=& g^2 \sba^2 \frac{\mw^2\epsilon_{+}\cdot\epsilon_{-} +\epsilon_{+}\cdot k_2\,\epsilon_{-}\cdot k_1}{u-\mw^2} \,, \nn\\
 \amp^{\rm 2HDM}_{\rm heavy}  \vert_c &=& \frac{g^2}{2} \,\epsilon_{+}\cdot\epsilon_{-}  \,.
\label{ampWWtohh-2HDM-matching}
\eear
By comparing first, the $t$- and $u$-channels of  \eqref{ampWWtohh-EChL} with those in \eqref{ampWWtohh-2HDM-matching},  we confirm the previous solution for $a$ in \eqref{matching-a}. 
Second,  by comparing the contact channel of \eqref{ampWWtohh-EChL} with the $s$-independent contribution of the $s$-channel and the contact channel of \eqref{ampWWtohh-2HDM-matching}, we arrive to:
\be
b = 1-\Delta b= 1 + \cba^2 (1 - 2\cba^2 + 2\cba\sba\cot2\beta)\,.
\label{matching-b}
\ee
Finally,  by plugging the previous solution for $a$ in \eqref{matching-a}  in the $s$-channel of \eqref{ampWWtohh-EChL} and comparing it with the $s$-dependent contribution of the $s$-channel of \eqref{ampWWtohh-2HDM-matching},  we find:
\be
\kappa_3 = 1- \Delta \kappa_3= \sba(1+2\cba^2) +\cba^2\left( -2\sba\mbarh +2\cba\cot2\beta \left(1-\mbarh\right) \right) \,.
\label{matching-kappa3}
\ee

{\bf 3) $ZZ \to hh$} 

The procedure here is similar than in the previous $WW$ scattering.  From \eqref{ampZZtohh-2HDM} we get the following  large heavy mass expansion results for the 2HDM amplitude,  presented again by channels:
\bear
 \amp^{\rm 2HDM}_{\rm heavy} \vert_s &=& \frac{g^2}{\cw^2} \left(\frac{3\sba}{s-\mh^2}\left(\sba\left(1+2\cba^2\right) \frac{\mh^2}{2} -\sba\cba^2\mbarc +\cba^3 \cot 2\beta \left(\mh^2-\mbarc\right)\right) \right.  \nn\\
 &&\left.\hspace{2mm}-\frac{\cba^2}{2} \left( -2 \cba \sba \cot 2\beta  +2\cba^2-1 \right)\right) \epsilon_{1}\cdot\epsilon_{2} \,, \nn\\
 \amp^{\rm 2HDM}_{\rm heavy}  \vert_t &=& \frac{g^2}{\cw^2} \sba^2 \frac{\mw^2\epsilon_{1}\cdot\epsilon_{2} +\epsilon_{1}\cdot k_1\,\epsilon_{2}\cdot k_2}{t-\mw^2} \,, \nn\\
 \amp^{\rm 2HDM}_{\rm heavy}  \vert_u &=& \frac{g^2}{\cw^2} \sba^2 \frac{\mw^2\epsilon_{1}\cdot\epsilon_{2} +\epsilon_{1}\cdot k_2\,\epsilon_{2}\cdot k_1}{u-\mw^2} \,, \nn\\
 \amp^{\rm 2HDM}_{\rm heavy}  \vert_c &=& \frac{g^2}{2\cw^2} \,\epsilon_{1}\cdot\epsilon_{2}  \,.
\label{ampZZtohh-2HDM-matching}
\eear
By comparing the  amplitudes from the HEFT in \eqref{ampZZtohh-EChL} with the previous 2HDM results, and by solving the matching equations in this $ZZ\to hh$ case,  we arrive at the same results  of \eqrefs{matching-a}{matching-kappa3}. Therefore, this channel serves as a cross-check of the previous solutions. 

{\bf 4) $hh \to hh$} 

For the case of  $hh\to hh$ scattering,  the results from \eqref{amphhtohh-2HDM} in the large heavy mass limit are given by  the following expressions,  also presented by channels:
\bear
 \amp^{\rm 2HDM}_{\rm heavy} \vert_s &=& -\frac{36}{\vev^2(s-\mh^2)}\left(\sba\left(1+2\cba^2\right) \frac{\mh^2}{2} -\sba\cba^2\mbarc +\cba^3 \cot 2\beta \left(\mh^2-\mbarc\right)\right)^2  \nn\\
 && +\frac{\mHeavy^2}{\vev^2}\cba^2\left( 2\cba^2-1-2\cba\sba\cot2\beta \right)^2  \nn\\
 && +\frac{2\cba}{\vev^2}\left( 2\cba^2-1-2\cba\sba\cot2\beta \right)\left( s\frac{\cba}{2}\left( 2\cba^2-1-2\cba\sba\cot2\beta \right) \right.  \nn\\
 &&\left. \hspace{5mm}+\mh^2 +\cba^2\left(-4\sba^2\mbarc +4 \cba^2 \cot^2 2\beta \left(-\mbarc+\cba^2\mh^2\right) \right.\right.  \nn\\
 &&\left.\left.\hspace{5mm}+4 \cba\sba \cot 2\beta \left(-2 \mbarc+\left(2\cba^2+1\right) \mh^2\right) -4 \cba^4\mh^2 +3\mh^2 \right) \right) \,, \nn\\
 \amp^{\rm  2HDM}_{\rm heavy} \vert_t &=& \amp^{\rm  2HDM}_{\rm heavy} \vert_s \quad {\rm with\,\,}s\to t \,, \nn\\
 \amp^{\rm 2HDM}_{\rm heavy} \vert_u &=& \amp^{\rm  2HDM}_{\rm heavy}  \vert_s \quad {\rm with\,\,}s\to u \,, \nn\\
 \amp^{\rm  2HDM}_{\rm heavy} \vert_c &=& -3\frac{\mHeavy^2}{\vev^2}\cba^2\left( 2\cba^2-1-2\cba\sba\cot2\beta \right)^2  \nn\\
 &&  -\frac{3}{\vev^2}\left( 4\cba^3\sba \cot2\beta \left((1+2\cba^2)\mh^2 - 2\mbarc\right) \right.  \nn\\
  && \left. - (-1 + \cba^2) \left((1+ 2 \cba^2)^2\mh^2 - 4\cba^2 \mbarc\right)  \right.  \nn\\
  && \left. + 4 \cot^2 2\beta \left(\cba^6 \mh^2 - \cba^4 \mbarc\right) \right) \,.
\label{amphhtohh-2HDM-matching}
\eear
Notice the $\mHeavy^2$ dependence in the separate contributions from the various channels.  In this particular case with scalar particles in all the external legs the result is not organized in different Lorentz structures that can be compared in solving the matching.  Thus,  the matching cannot be analyzed separately by channels,  and it must  be done instead by using the total sum, i.e the matching quantity is the total amplitude.  Then, when adding the contributions from all the channels,  one finds that the potentially growing terms with the heavy mass of $\mO(\mHeavy^2)$ in the separate contributions are totally canceled in the sum. We have checked this cancellation explicitly in the total amplitude.  Thus,  the leading dependence on the heavy mass of the total amplitude,  after the integration of the heavy modes,  is again of ${\cal O}(m_{\rm EW}^2/m_{\rm heavy}^2)^0$,  providing non-decoupling contributions that are constant with $m_{\rm heavy}$.    

In addition, the growing energy behavior in the contributions from the separate $s$, $t$ and $u$ channels also disappear in the total amplitude after summing over them using the relation $s+t+u=4\mh^2$.
The resulting total amplitude has contributions decreasing with the  $s$, $t$ and $u$ variables,  and comparing them with the corresponding HEFT contributions of \eqref{amphhtohh-EChL},  we arrive to the same solution for $\kappa_3$ as obtained previously in \eqref{matching-kappa3}.  Thus,  this part serves as  a cross-check for $\kappa_3$. 
Finally,  by solving the matching equation looking at the contribution to the total amplitude that is constant in energy,  one gets  the following result for $\kappa_4$:
\bear
\kappa_4 &=1-\Delta \kappa_4= & 1 +\frac{\cba^2}{3}\left( -7+64\cba^2-76\cba^4 +12(1-6\cba^2+6\cba^4)\mbarh \right.  \nn\\
&&\left. \hspace{7mm}+4\cba\sba\cot 2\beta \left(-13+38\cba^2-3(-5+12\cba)\mbarh\right) \right.  \nn\\
&&\left. \hspace{7mm}+4\cba^2\cot^2 2\beta \left(3\cba^2-16\sba^2+3(-1+6\sba^2)\mbarh\right) \right)\,.
\label{matching-kappa4}
\eear

{\bf 5) $h \to \gamma \gamma$ and $h \to \gamma Z$}

Finally, for the Higgs boson decays $h \to \gamma \gamma$ and $h \to \gamma Z$,  the results for the one-loop amplitude summarizing the heavy mass limit of the charged Higgs boson loops are obtained  from \eqref{hAAloop-Hp} and \eqref{hAZloop-Hp},  and by taking into account the large mass behavior of the functions $f(r)$ and $g(r)$ given in  \eqref{fyg-larger}. The results for the functions defining those amplitudes are the following:
\bear
F_{h\gamma\gamma}^{H^\pm} \vert_{\rm heavy} &=& -\frac{g^2\sw^2\sba}{48\pi^2}\,, \nn\\
F_{h\gamma Z}^{H^\pm} \vert_{\rm heavy} &=& -\frac{g^2\sw(2\cw^2-1)\sba}{96\cw\pi^2}\,.
\label{amploop-2HDM-matching}
\eear
Plugging these results  in to the respective matching equations,  we finally find the solutions for the HEFT coefficients: 
\bear
a_{h\gamma\gamma} &=& -\frac{\sba}{48\pi^2} \,, \nn\\
a_{h\gamma Z} &=& -\frac{(2\cw^2-1)\sba}{96\cw^2\pi^2}\,.
\label{matching-loop}
\eear
In addition,   these decay channels also confirm the solution for $a$ given in \eqref{matching-a}. 
Notice that these two coefficients do not vanish in the alignment limit,  therefore they may have relevant phenomenological implications.  In particular,  the implications of the non-decoupling $H^\pm$ loops in BR$(h \to \gamma \gamma)$ have been recently analyzed in the context of the LHC physics in~\cite{Arco:2022jrt},  and they find sizable departures respect to the SM rates, even for large 
$m_{H^\pm}$ near the TeV.   Other non-decoupling effects in the 2HDM from loops with heavy $H^\pm$  have also been found in flavor changing Higgs decays, $h \to b \bar s$ and $h\to s \bar b$,  in 
\cite{Arco:2023hmz}.  In both works,  the role of a large triple Higgs coupling $\lambda_{hH^+H^-}$ has been pointed out.

\subsection{HEFT coefficients from non-decoupling heavy Higgs bosons}
Finally,  we put together here all the analytical results for the HEFT coefficients found in the previous section by solving the full set of  matching equations. We provide these results in terms of the HEFT coefficients that define the BSM contributions from the 2HDM with respect to the SM.  These are the following (we add explicitly here the label 2HDM for completeness): 
\bear
\Delta a\vert_{\rm 2HDM} &=& 1 -\sba \,, \nn\\
\Delta b\vert_{\rm 2HDM} &=& - \cba^2 (1 - 2\cba^2 + 2\cba\sba\cot2\beta) \,, \nn\\
\Delta \kappa_3\vert_{\rm 2HDM}   &= & 1- \sba(1+2\cba^2) -\cba^2\left( -2\sba\mbarh +2\cba\cot2\beta \left(1-\mbarh\right) \right) \,, \nn\\
\Delta \kappa_4\vert_{\rm 2HDM}  &=& -\frac{\cba^2}{3}\left( -7+64\cba^2-76\cba^4 +12\left(1-6\cba^2+6\cba^4\right)\mbarh \right.  \nn\\
&&\left. \hspace{7mm}+4\cba\sba\cot 2\beta \left(-13+38\cba^2-3(-5+12\cba)\mbarh\right) \right.  \nn\\
&&\left. \hspace{7mm}+4\cba^2\cot^2 2\beta \left(3\cba^2-16\sba^2+3(-1+6\sba^2)\mbarh\right) \right) \,, \nn \\
a_{h\gamma\gamma}\vert_{\rm 2HDM}  &=& -\frac{\sba}{48\pi^2} \,, \nn\\
a_{h\gamma Z}\vert_{\rm 2HDM}  &=& -\frac{(2\cw^2-1)\sba}{96\cw^2\pi^2}\,.
\label{matching-full}
\eear
Some comments are in order.  First,  for shortness,  in the previous formulas  we have used again a compact form.  To get the explicit result in terms of the 2HDM input parameters,  the values of $ \sba$ ,  $\sb$,  $\cb$ and  $\cot2\beta$ given in \eqref{trigo} should be plugged in all these formulas.  Then, the first conclusion is that these HEFT coefficients,  capturing the non-decoupling effects from the heavy Higgs bosons in the 2HDM,  depend on the subset of input parameters given by $\cba$, $\tan \beta$,  $m_h$  and $m_{12}$.  Second,  the above results are valid for arbitrary $-1 \leq \cba \leq 1$, therefore they set the coefficient values for the generic scenario  with misalignment ($\cba \neq 0$).  Third,  in the case of an scenario with alignment, i.e.  with $\cba = 0$,  we get vanishing LO-HEFT $\Delta$'s.  More interestingly, 
we get non-vanishing values in this alignment limit for the NLO-HEFT coefficients 
$a_{h \gamma \gamma}$ and $a_{h\gamma Z}$.  Specifically,  we get:
\bear
\Delta a\vert_{\rm 2HDM}^{\rm al} &=& 0 \,, \nn\\
\Delta b\vert_{\rm 2HDM}^{\rm al} &=&0 \,, \nn\\
\Delta \kappa_3\vert_{\rm 2HDM}^{\rm al}   &= & 0 \,, \nn\\ 
\Delta \kappa_4\vert_{\rm 2HDM}^{\rm al}  &=& 0 \,, \nn\\
a_{h\gamma\gamma}\vert_{\rm 2HDM}^{\rm al}   &=& -\frac{1}{48\pi^2} \,, \nn\\
a_{h\gamma Z}\vert_{\rm 2HDM}^{\rm al}   &=& -\frac{2\cw^2-1}{96\cw^2\pi^2}\,.
\label{matching-align}
\eear
 
Fourth,  in the case of an scenario with quasi-alignment,  namely with small but not vanishing  $|\cba| \ll 1$,  we can approximate the above results in \eqref{matching-full} by doing an additional Taylor expansion in powers of the small parameter $\cba$ and keeping  just the leading term in this expansion,  which for the LO-HEFT $\Delta$'s  is of ${\cal O}(\cba^2)$.  Thus,  we get the following results for these $\Delta$'s in this quasi-alignment (qal) scenario:
\bear
\Delta a\vert_{\rm 2HDM}^{\rm qal} &=&  \frac{\cba^2}{2} \,,  \nn\\
\Delta b\vert_{\rm 2HDM}^{\rm qal} &=& -\cba^2 \,, \nn\\
\Delta \kappa_3\vert_{\rm 2HDM}^{\rm qal}   &= & -\cba^2\left(\frac{3}{2}-2\frac{m_{12}^2}{\mh^2}\frac{1+\tanb^2}{\tanb}\right) \,, \nn\\ 
\Delta \kappa_4\vert_{\rm 2HDM}^{\rm qal}  &=& \cba^2\left(\frac{7}{3}-4\frac{m_{12}^2}{\mh^2}\frac{1+\tanb^2}{\tanb}\right) \,.
\label{matching-quasialign}
\eear

Fifth,  we  remark that in the previous quasi-alignment limit we find some simple correlations among the LO-HEFT coefficients that we find interesting to comment:
\bear
\label{correlation-quasialign1}
\Delta a\vert_{\rm 2HDM}^{\rm qal} &=& -\frac{1}{2}  \Delta b\vert_{\rm 2HDM}^{\rm qal} \,,  \\
\Delta \kappa_3\vert_{\rm 2HDM}^{\rm qal} &=& -\frac{9}{14} \Delta \kappa_4\vert_{\rm 2HDM}^{\rm qal} 
 -\frac{4}{7}\cba^2\frac{m_{12}^2}{\mh^2}\frac{1+\tanb^2}{\tanb} \,.  
\label{correlation-quasialign2}
\eear
The first one is independent on $m_{12}$ and $\tan\beta$.  Notice the different correlation between  these two 
$\Delta$'s here in the 2HDM and in the SMEFT which was commented at the end of~\secref{section-EChL}.  For the result regarding the $\kappa$'s  in \eqref{correlation-quasialign2} the correlation gets further simplified  if $m_{12}=0$: 
\bear
\Delta \kappa_3\vert_{\rm 2HDM}^{\rm qal} &=
& -\frac{9}{14} \Delta \kappa_4\vert_{\rm 2HDM}^{\rm qal}  \,\,\,\, (m_{12}=0)    \,.
\label{k3k4-m12i0}
\eear
On the other hand,  one can also write together  $\Delta \kappa_3\vert_{\rm 2HDM}^{\rm qal}$ and  $\Delta \kappa_4\vert_{\rm 2HDM}^{\rm qal}$ of \eqref{matching-quasialign} in the following alternative form:
\bear
2 \Delta \kappa_3\vert_{\rm 2HDM}^{\rm qal} + \Delta \kappa_4\vert_{\rm 2HDM}^{\rm qal}&=
& -\frac{2}{3} \cba^2\,,
\label{correlation-quasialign3} 
\eear
which is interesting because it is independent on the value of $m_{12}$  and $\tan\beta$.  Then,  from
 \eqref{matching-quasialign} and \eqref{correlation-quasialign3} it is immediate to check that the following relation among the four $\Delta$'s holds in the quasi-alignment case: 
 \bear
 \label{correlation-quasialign4} 
 2 \Delta \kappa_3\vert_{\rm 2HDM}^{\rm qal} + \Delta \kappa_4\vert_{\rm 2HDM}^{\rm qal}&=&
 \frac{2}{3} \Delta b\vert_{\rm 2HDM}^{\rm qal} = -\frac{4}{3} \Delta a\vert_{\rm 2HDM}^{\rm qal} \,.
 \eear

Our final comment refers to the comparison of our results for the HEFT coefficients with those found in~\cite{Dawson:2023ebe}.  They do not consider one-loop generated coefficients,  and work close to the alignment limit,  thus we can only compare our simplified results in the quasi-alignment limit in \eqref{matching-quasialign} with their Eq.~(23).  Except for $\Delta a$,  where we agree, the rest of $\Delta$'s are clearly different.  In fact, they express the results for the HEFT-coefficients in terms of a common heavy mass $\Lambda$ and the splittings among the heavy Higgs boson masses which they assume to be of ${\cal O}(v)$.  They also assume  a value for $\cba$ of  ${\cal O}(v^2/\Lambda^2)$.  None of these assumptions are done in the present work,  since we deal with generic input $\cba$ and generic mass splittings,  so these different results are not surprising.  Our interpretation of these differences is that they follow different paths to move through the 2HDM parameter space in the heavy mass limit.  In particular, they `freeze' the values of the triple Higgs couplings
$\lambda_{\rm x}$ by their perturbativity requirement  whereas we do not.  Thus, they do not get non-decoupling effects from the heavy Higgs bosons, whereas we do.

\section{Numerical Results}
\label{section-numerical}

\begin{figure}[t!]
	\centering
	\begin{tabular}{cc}
		\includegraphics[width=0.36\textwidth]{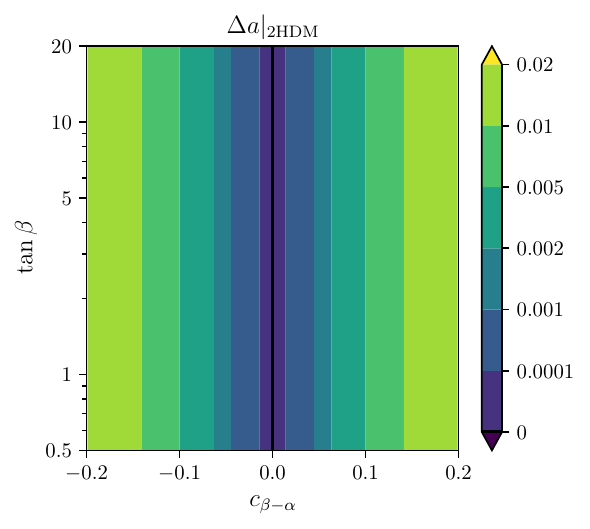} & \includegraphics[width=0.36\textwidth]{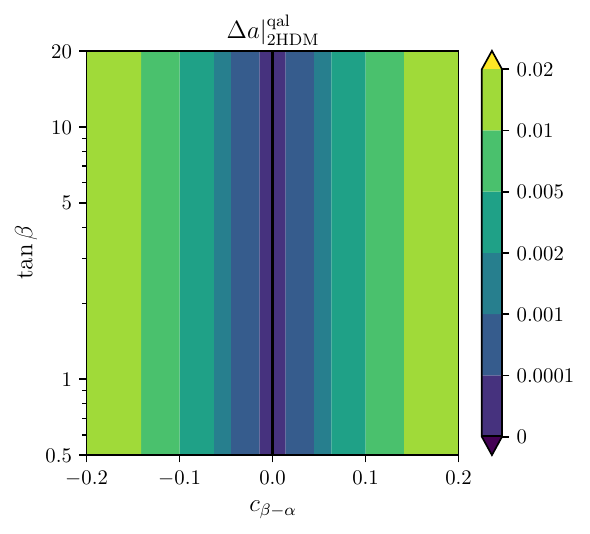} \\
		\includegraphics[width=0.36\textwidth]{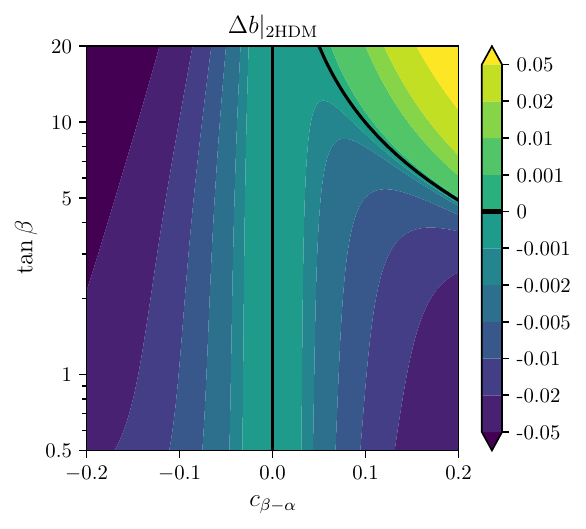} & \includegraphics[width=0.36\textwidth]{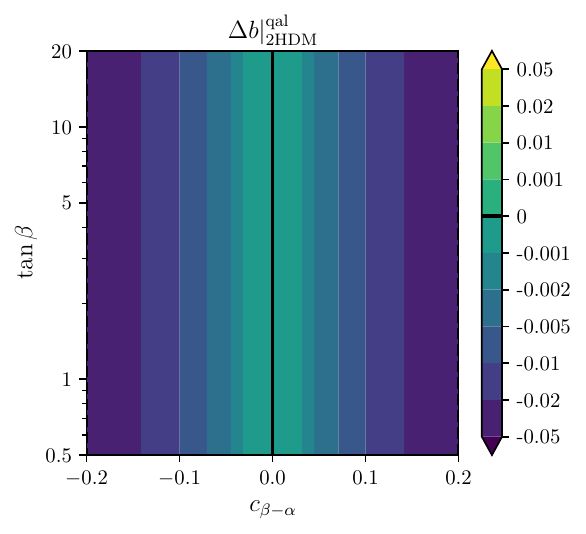}\\
		\includegraphics[width=0.36\textwidth]{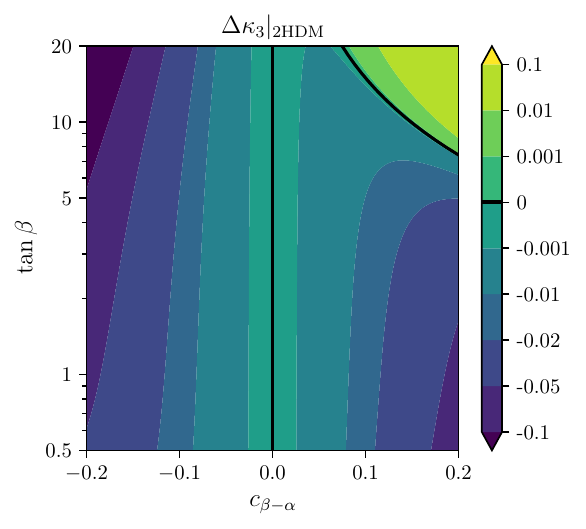} & \includegraphics[width=0.36\textwidth]{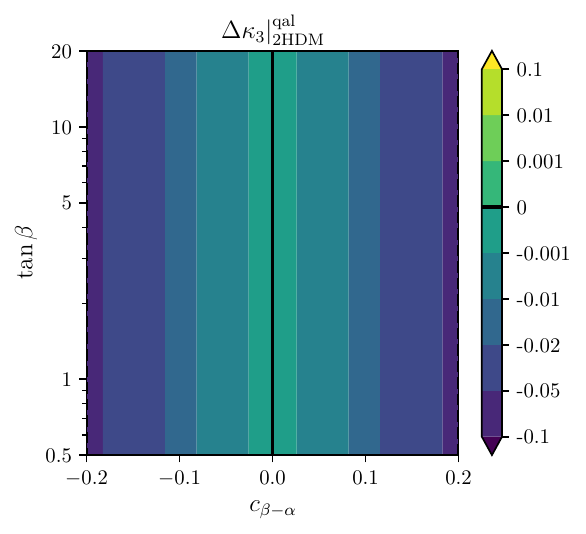}\\
		\includegraphics[width=0.36\textwidth]{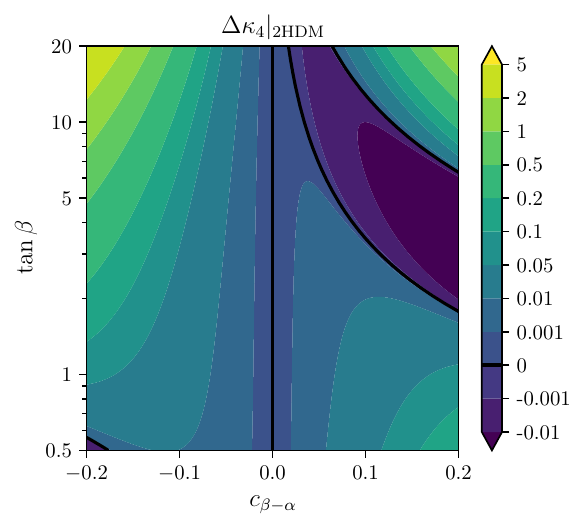} & \includegraphics[width=0.36\textwidth]{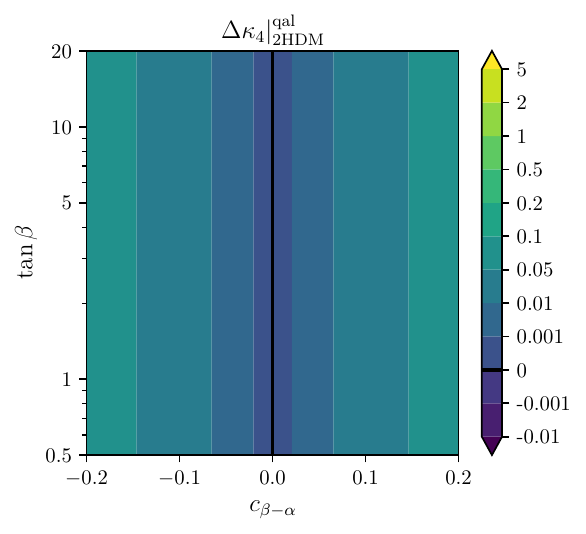}
	\end{tabular}
	\caption{LO-HEFT parameters, $\Delta a$,  $\Delta b$,   $\Delta \kappa_3$,  $\Delta \kappa_4$ from 2HDM: Contours in the $(\cba, \tan \beta)$ plane for  $m_{12}=0$.  Misalignment (left plots),  using 
	\eqref{matching-full}, versus quasi-alignment (right plots),  using \eqref{matching-quasialign}. }
	\label{plots-full-qa-m120}
\end{figure}

\begin{figure}[t!]
	\centering
	\begin{tabular}{cc}
		\includegraphics[width=0.35\textwidth]{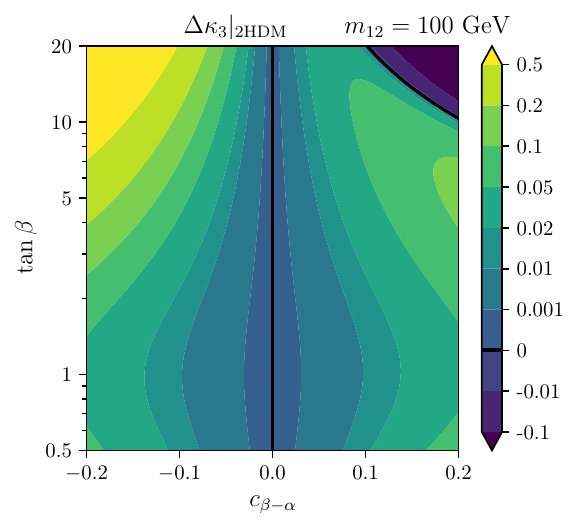} & \includegraphics[width=0.35\textwidth]{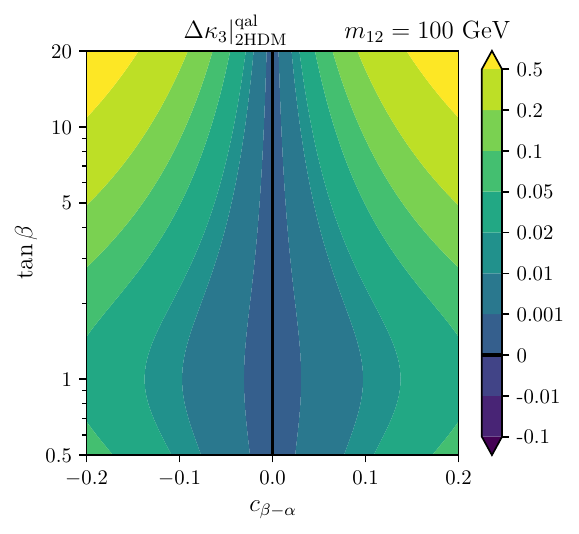}\\
		\includegraphics[width=0.35\textwidth]{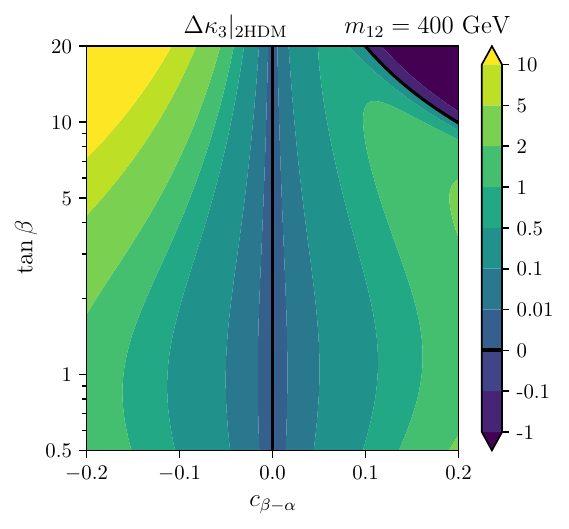} & \includegraphics[width=0.35\textwidth]{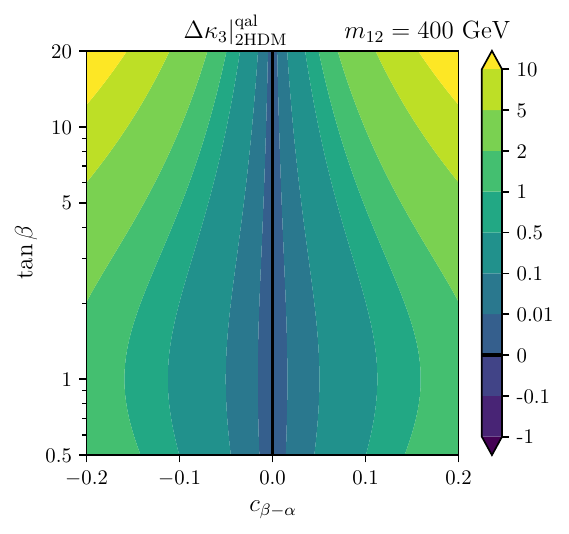}\\
		\includegraphics[width=0.35\textwidth]{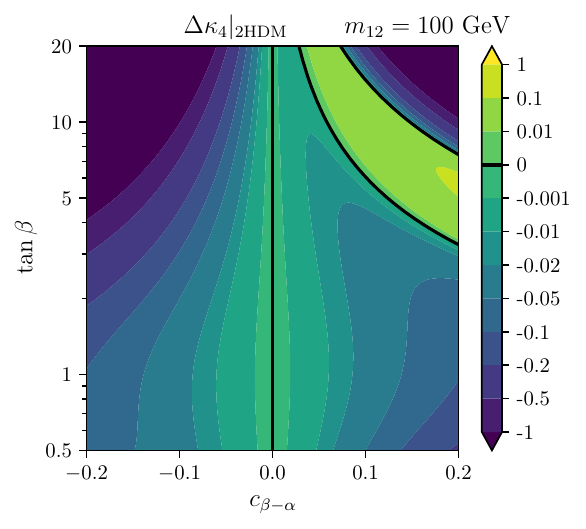} & \includegraphics[width=0.35\textwidth]{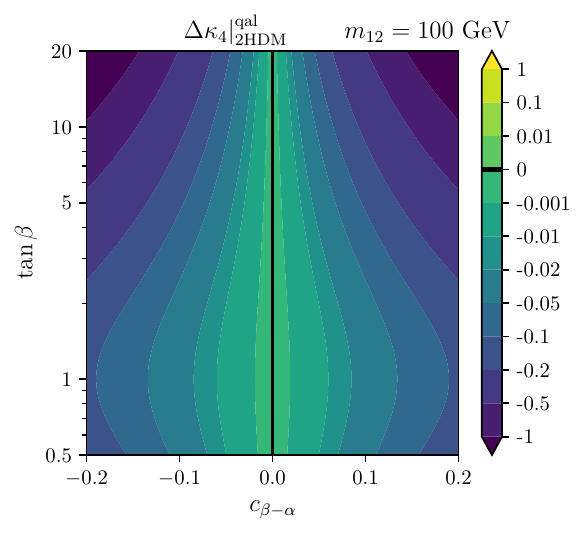}\\
		\includegraphics[width=0.35\textwidth]{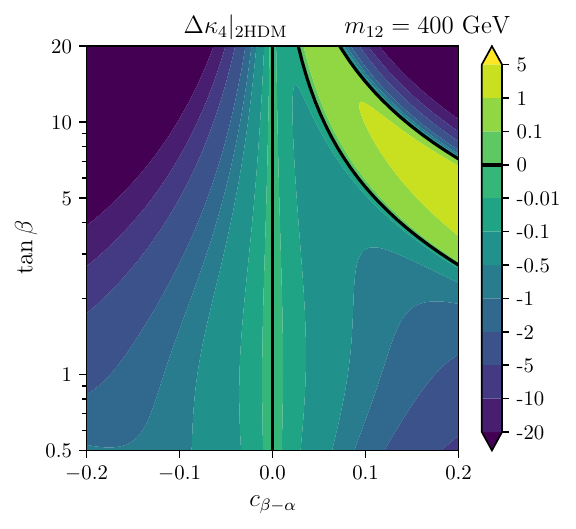} & \includegraphics[width=0.35\textwidth]{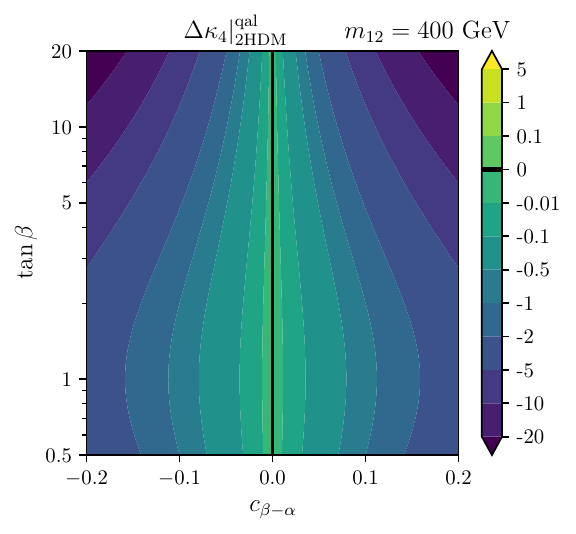}
	\end{tabular}
	\caption{LO-HEFT parameters,  $\Delta \kappa_3$,  $\Delta \kappa_4$,  from 2HDM:  Contours in the $(\cba, \tan \beta)$ plane for  $m_{12} \neq 0$.  Misalignment (left plots), using 
	\eqref{matching-full},  versus quasi-alignment (right plots), using \eqref{matching-quasialign}.}
	\label{plots-full-qa-m12no0}
\end{figure}

In this section we present our numerical results for the HEFT coefficients in \eqref{matching-full} in terms of the 2HDM input parameters. 
First,  we comment on the one-loop generated coefficients $a_{h\gamma\gamma}\vert_{\rm 2HDM}$ and
$a_{h\gamma Z}\vert_{\rm 2HDM}$.  Since they only depend on the value of  $\sba$,  and this lies in the interval $0\leq\sba\leq 1$, then the predicted coefficients in \eqref{matching-full} fulfill:
\bear
-\frac{1}{48\pi^2} \leq &
a_{h\gamma\gamma}\vert_{\rm 2HDM} & \leq 0  \,, \\
  -\frac{2\cw^2-1}{96\cw^2\pi^2}   \leq &  a_{h\gamma Z}\vert_{\rm 2HDM} & \leq 0  \,.
\eear
Then,  they are numerically small quantities, as it is expected since they are one-loop generated coefficients. 
In particular,  the values reached for alignment are the following:
\bear
a_{h\gamma\gamma}\vert_{\rm 2HDM}^{\rm al} &=& -0.00211  \,, \\
a_{h\gamma Z}\vert_{\rm 2HDM}^{\rm al}  &=& -0.000944 \,.
\eear 

Next,  we comment on the LO-HEFT coefficients.  We start with the simplest case of $m_{12}=0$.  The numerical predictions for $\Delta a\vert_{\rm 2HDM}$,   $\Delta b\vert_{\rm 2HDM}$,  $\Delta \kappa_3\vert_{\rm 2HDM}$ and  $\Delta \kappa_4\vert_{\rm 2HDM}$ in this $m_{12}=0$ case depend just on  $\cba$, $\tan \beta$.  Their predictions in terms of these two parameters are presented in \figref{plots-full-qa-m120},  as contours lines in the $(\cba, \tan \beta)$ plane.  The predictions in the misalignment scenario are collected in the left plots and the predictions in the quasi-alignment scenario in the right plots.  
The black solid lines in these plots represent the predictions where the $\Delta$'s are zero.
This always coincides with the alignment limit, but the $\Delta$'s can also vanish in other regions with concrete configurations of the 2HDM input parameters.
The intervals displayed in the axes of these plots are chosen to roughly cover the allowed experimental values.  In particular,  the interval explored in $\cba$ is shorten to  $|\cba|<0.2$ and  the one in $\tan\beta$ is shorten to $0.5 <\tan \beta <20$.    The numerical values predicted in those reduced intervals range roughly as follows: $0<\Delta a\vert_{\rm 2HDM}<0.02$,  $-0.2<\Delta b\vert_{\rm 2HDM}<0.12$,  
$-0.22<\Delta \kappa_3\vert_{\rm 2HDM}<0.11$ and $-0.05<\Delta \kappa_4\vert_{\rm 2HDM}<4.5$.  Their maximum values are reached for the largest considered $|\cba|$ values  and the largest considered $\tan \beta$ values (except for $\Delta a$ that is independent on $ \tan \beta$).  Regarding the comparison between the misalignment and quasi-alignment results, we see that the predictions for $\Delta a\vert_{\rm 2HDM}$ (left) versus those for $ \Delta a\vert_{\rm 2HDM}^{\rm qal}$ (right) look very similar.  We also see that the predictions for the other $\Delta$'s look also very similar in the low region of $\tan \beta$.  The largest differences among misalignment and quasi-alignment occur in the upper right corner for $\Delta b$ and $\Delta{\kappa_3}$ and in the upper left corner for $\Delta{\kappa_4}$.  In any case we can conclude that the simple formulas in \eqref{matching-quasialign} (applied for $m_{12}=0$)  provide a very good approximation to the full result of \eqref{matching-full} in the low  $\tan \beta <2$ region.

\begin{figure}[t!]
	\centering
	\begin{tabular}{cc}
		\includegraphics[width=0.48\textwidth]{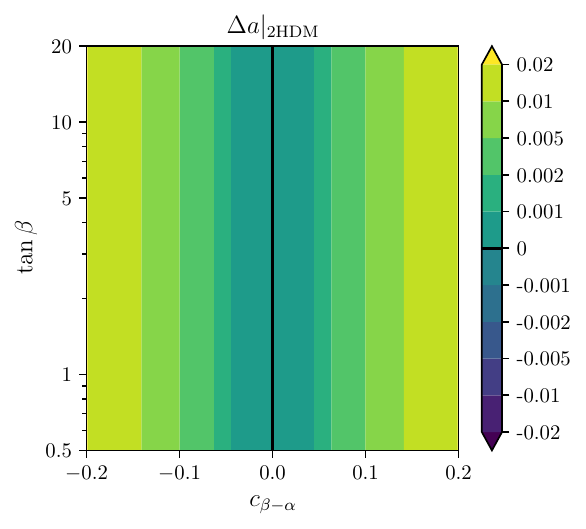} & \includegraphics[width=0.48\textwidth]{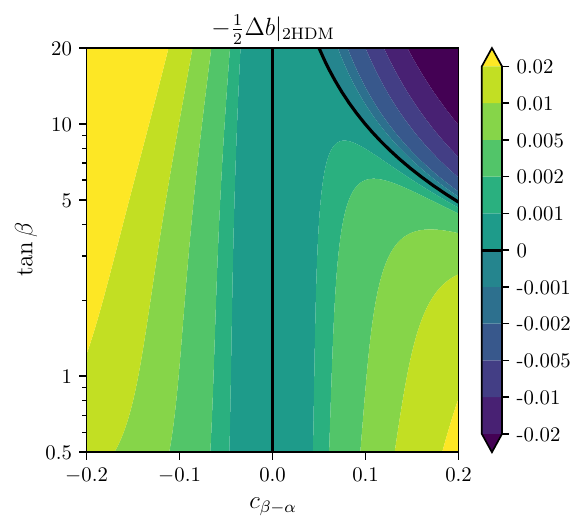}
	\end{tabular}
	\caption{Correlations between  LO-HEFT coefficients from the 2HDM,  $\Delta a\vert_{\rm 2HDM}$ and $-\frac{1}{2}\Delta b\vert_{\rm 2HDM}$ in the misalignment case,  using \eqref{matching-full}. }
	\label{plot-DavsDb}
\end{figure}

\begin{figure}[t!]
	\centering
	\begin{tabular}{cc}
		\includegraphics[width=0.48\textwidth]{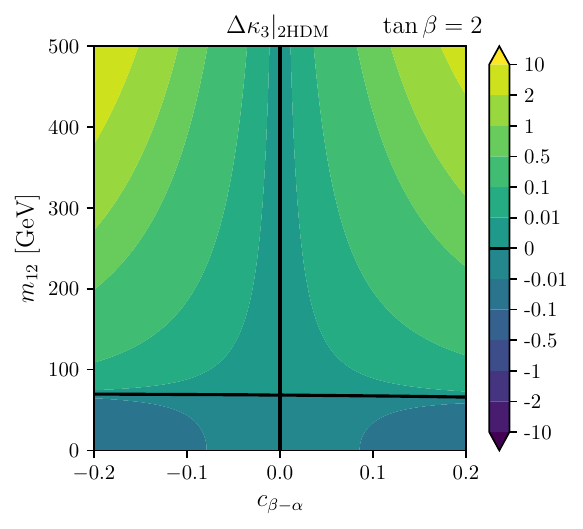} & \includegraphics[width=0.48\textwidth]{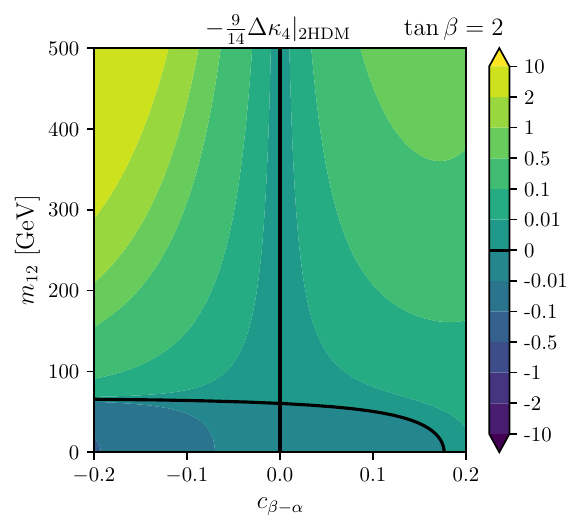}  \\
		\includegraphics[width=0.48\textwidth]{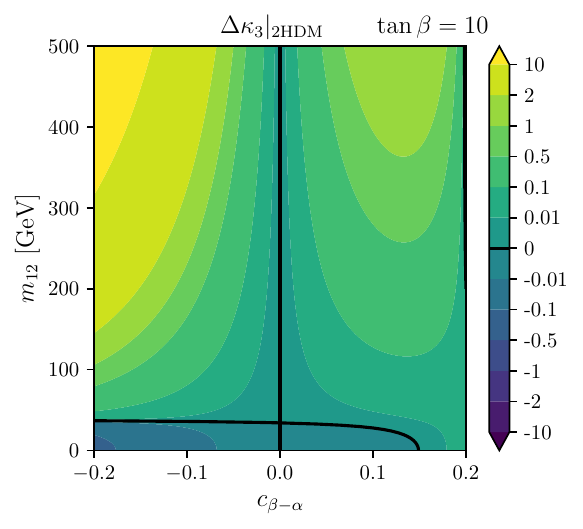} & \includegraphics[width=0.48\textwidth]{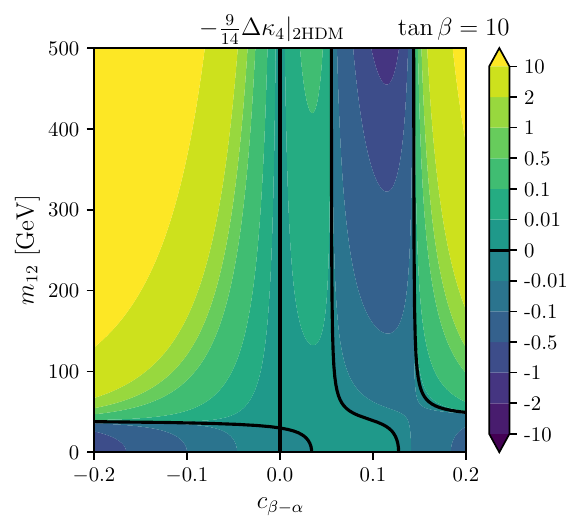}
	\end{tabular}
	\caption{Correlations between LO-HEFT coefficients from the 2HDM 
		$\Delta \kappa_3\vert_{\rm 2HDM}$ and $-\frac{9}{14}\Delta \kappa_4\vert_{\rm 2HDM}$  in the misalignment case, using \eqref{matching-full}. }
	\label{plot-Dk3vsDk4}
\end{figure}

The values of $\Delta \kappa_3$ and $\Delta \kappa_4$ for the $m_{12} \neq 0$ case are explored in \figref{plots-full-qa-m12no0}.  We consider two different input values of $m_{12}=100$ GeV  and $m_{12}=400$ GeV.  The same intervals in the axes as before are considered in these contours in the $(\cba, \tan \beta)$ plane for the $m_{12} \neq 0$ case.  Looking at the misalignment plots (left) we see that the size  of
$\Delta \kappa_3$ increases for larger $m_{12}$ reaching values of up to 10 in the upper left corner (with $\tan \beta$ close to 20)  of the second plot for $m_{12}=400$ GeV.  The size of  $\Delta \kappa_4$ also reaches the largest values of about 5 for the larger $m_{12}=400$ GeV case,  but it happens at lower values of $\tan \beta$ below 10.
In addition, $\Delta \kappa_4$ can also reach large, but negative, values below~-20 in the regions of large $\tan\beta$ and far from the alignment limit in both cases where $m_{12}=100,\,400$~GeV.
Comparing the plots on the left (misalignment) with those in the right  (quasi-alignment) we find again that these later provide a reasonable approximation in the low $\tan \beta$ region,  roughly below 5 for $\Delta \kappa_3$ and below 2 for $\Delta \kappa_4$.  

Finally, we study numerically the interesting correlations found among the HEFT-coefficients,  $\Delta a$ versus $\Delta b$ and $\Delta \kappa_3$ versus $\Delta \kappa_4$ in 
\figref{plot-DavsDb} and \figref{plot-Dk3vsDk4},  respectively.
Notice that we have chosen to plot in \figref{plot-DavsDb} and \figref{plot-Dk3vsDk4} the quantities $\Delta a\vert_{\rm 2HDM}$ versus  $-\frac{1}{2} \Delta b\vert_{\rm 2HDM}$ and $\Delta \kappa_3\vert_{\rm 2HDM}$ versus $-\frac{9}{14} \Delta \kappa_4\vert_{\rm 2HDM}$  motivated by the particular combinations appearing in the quasi-alignment results of \eqref{correlation-quasialign1} and \eqref{correlation-quasialign2}, respectively. 
These are all studied for the general case with misalignment, namely, we use the full formulas in \eqref{matching-full}.  One can see in \figref{plot-DavsDb} that a clear correlation between $\Delta a\vert_{\rm 2HDM}$ and $-\frac{1}{2} \Delta b\vert_{\rm 2HDM}$ is manifested for $\tan \beta <5$,  and this correlation is well represented by our approximate results of the quasi-alignment scenario.  Regarding  $\Delta \kappa_3$ and $\Delta \kappa_4$,  we see in the upper plots in \figref{plot-Dk3vsDk4} that for low $\tan \beta =2$ a clear correlation between $\Delta \kappa_3\vert_{\rm 2HDM}$ and 
$-\frac{9}{14} \Delta \kappa_4\vert_{\rm 2HDM}$ is manifested for  $m_{12}<300$ GeV.  Increasing $\tan \beta$ worsen this correlation.  From the two lower plots,  for $\tan \beta=10$,  we see that this correlation manifest only in the very narrow region with very small $\cba$ values close to alignment.  Therefore this correlation is well represented by our aproximated formulas of the quasi-alignment scenario for sufficiently small $\cba$ values,  namely,  for
$|\cba| < 0.05$. 

\section{Conclusions}
\label{section-conclus}
In this work we have computed the most relevant non-decoupling effects from the BSM Higgs bosons within the 2HDM,  $H$, $A$ and $H^\pm$.  The light Higgs boson $h$ is assumed here to be the one observed experimentally with a mass $m_h\sim 125$ GeV.  Our simple hypothesis for the masses of the BSM Higgs bosons is that they are very heavy compared to the EW masses,  $m_Z$, $m_W$, $m_f$,  $m_h$, $v$ and $m_{12}$.  We have worked within the framework of EFTs and more concretely we have assumed the HEFT to be the proper EFT to describe the low energy effects from the 2HDM  heavy Higgs bosons.  We focus here on just the bosonic sector.  Specifically,   we have found that the low energy effects that result from the integration of the heavy  Higgs boson modes $H$, $A$ and $H^\pm$ can be collected into  a set of HEFT coefficients which turn out not to be suppressed by  inverse powers of the heavy masses but instead they are constant with these masses. 
Those values constant with $m_{\rm heavy}$ summarize the non-decoupling effects from the heavy Higgs bosons.

Our approach to compute such non-decoupling effects is by solving the matching between the 2HDM and the HEFT at low energies compared to the heavy masses.  Instead of the usual matching at the Lagrangian level,  we impose here a more physical matching which requires the equality between the amplitudes predicted by the 2HDM in the heavy mass limit of the BSM Higgs bosons with those predicted by the HEFT.  
Furthermore, we do this matching at the amplitude level by considering specific processes involving the light Higgs boson in the external legs  that include scattering and decays.  Concretely, we have studied and solved the matching between the 2HDM and the HEFT amplitudes for the following seven processes:  $h\to WW^*\to Wf\bar{f'}$,  $h\to ZZ^*\to Zf\bar{f}$,  $W^+W^- \to hh$,  $ZZ \to hh$,  $hh \to hh$,  $h \to \gamma \gamma$ and $h \to \gamma Z$.  All amplitudes have been computed in a covariant $R_\xi$ gauge to get control on the gauge invariance of the results. The amplitudes for the five first processes have been computed at the tree level in the HEFT,  2HDM and also the SM for comparison.  The amplitudes of the two last decays have been computed to one-loop level in the three models.  We have shown that the
expansion of the 2HDM amplitudes in inverse powers of the heavy masses provides convergent results for the HEFT coefficients in the heavy mass limit.   In addition, we have identified  the triple couplings of the light Higgs with the heavy Higgs bosons as being the responsible for the non-decoupling effects from the heavy Higgs bosons in the amplitudes. 

The non-decoupling effects found here are summarized in the values of the HEFT coefficients collected in \eqref{matching-full} which have been given in terms of the input 2HDM parameters.  These input parameters have been chosen in this work to be $m_h$,  $m_H$,  $m_A$, 
$m_{H^\pm}$, $v$,  $\cba$, $\tan \beta$ and $m_{12}$.  In fact,  the analytical results found here for the HEFT coefficients turn out to depend on just a subset of them.  Concretely,  $\Delta a$,  $a_{h \gamma \gamma}$ and $a_{h \gamma Z}$ are given in terms of just $\cba$.  $\Delta b$ is given in terms of $\cba$ and $\tan \beta$,   and $\Delta \kappa_3$ and $\Delta \kappa_4$ are given on terms of $\cba$,  $\tan \beta$ and $m_{12}$.   We wish to emphasize that these results are valid for a generic value of $\cba$, i.e,  they apply for the generic misalignment case.  We have also provided their analytical values in the simpler cases of alignment with $\cba=0$,  and of quasi-alignment with $\cba \ll 1$,  summarized in \eqref{matching-align} and \eqref{matching-quasialign} respectively.  In looking at those solutions from the matching we have detected some correlations among the HEFT coefficients, which can be of much interest in the future colliders searches of BSM physics. 

Finally, we have also  explored numerically the values of the HEFT coefficients  as a function of the 2HDM input parameters.  For the considered intervals in the relevant 2HDM input parameters, which are roughly allowed by the present constraints,  we find values of $0<\Delta a\vert_{\rm 2HDM}<0.02$,  $-0.2<\Delta b\vert_{\rm 2HDM}<0.12$,  
$-0.22<\Delta \kappa_3\vert_{\rm 2HDM}<0.11$ and $-0.05<\Delta \kappa_4\vert_{\rm 2HDM}<4.5$ in the simplest case of $m_{12}=0$.

The correlations among the HEFT coefficients have also been explored numerically.  We have found that a clear correlation between $\Delta a\vert_{\rm 2HDM}$ and $-\frac{1}{2} \Delta b\vert_{\rm 2HDM}$ is manifested for $\tan \beta <5$,  and this correlation is well represented by our approximate results of the quasi-alignment scenario.  Regarding the $\kappa$'s,  we have found that for low $\tan \beta =2$ a clear correlation between $\Delta \kappa_3\vert_{\rm 2HDM}$ and 
$-\frac{9}{14} \Delta \kappa_4\vert_{\rm 2HDM}$ is manifested for  $m_{12}<300$ GeV.  Increasing $\tan \beta$ worsen this correlation.  For instance,  for $\tan \beta=10$,  we see that this correlation manifest only in the very narrow region with very small $\cba$ values close to alignment.  Therefore this correlation is well represented by our aproximated formulas of the quasi-alignment scenario for sufficiently small $\cba$ values,  namely,  for
$|\cba| < 0.05$. 

All in all, we conclude that the non-decoupling effects found in this work could serve in the future as a guide to look for indirect hints from the 2HDM heavy Higgs bosons,  even if they are too heavy to be produced directly at colliders.

\section*{Acknowledgments}

F.A., D.D. and M.J.H. acknowledge financial support from the grant IFT Centro de Excelencia Severo Ochoa CEX2020-001007-S funded by MCIN/AEI/10.13039/501100011033, from the Spanish “Agencia Estatal de Investigación” (AEI) and the EU “Fondo Europeo de Desarrollo Regional” (FEDER) through the project PID2019-108892RB-I00 funded by MCIN/AEI/10.13039/501100011033, and from the European Union’s Horizon 2020 research and innovation programme under the Marie Sklodowska-Curie grant agreement No 860881-HIDDeN.
The work of R.M. is also supported by CONICET and ANPCyT under projects PICT 2017-2751, PICT 2018-03682 and PICT-2021-I-INVI-00374.
The work of F.A. is also supported by the grant PID2019-110058GB-C21 funded by

\hspace{-8mm}MCIN/AEI/10.13039/501100011033 and by ”ERDF A way of making Europe” and by the Spanish Ministry of Science and Innovation via an FPU grant No FPU18/06634.
\newpage

\section*{Appendices}
\appendix

\section{Relevant Feynman rules}
\label{App-Frules}

In this section we provide in \tabref{relevant-FR} all relevant Feynman rules for the computation of the scattering and decay amplitudes discussed in this work. We only include those interactions among fields that are external legs in the considered processes.  The three models in consideration, SM, HEFT and 2HDM, are showed for comparison.
The relevant Feynman rules for Higgs boson scalar interactions in the 2HDM can also be found in \eqrefs{eq-hhh}{eq-hhhh}.

\begin{table}[H]
\begin{tabular}{ >{\centering\arraybackslash}m{45mm}| >{\arraybackslash}c| >{\arraybackslash}c| >{\arraybackslash}c}
Interaction & SM & HEFT & 2HDM \\
\hline 
\includegraphics[width=31mm]{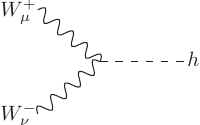} & $\frac{2i\mw^2}{\vev}g^{\mu\nu}$ & $\frac{2i \mw^2}{\vev}a g^{\mu\nu}$ & $\frac{2i \mw^2}{\vev}\sba g^{\mu\nu}$  \\
\includegraphics[width=31mm]{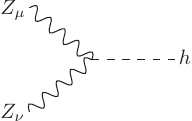} & $\frac{2i\mz^2}{\vev}g^{\mu\nu}$ & $\frac{2i \mz^2}{\vev}a g^{\mu\nu}$ & $\frac{2i \mz^2}{\vev}\sba g^{\mu\nu}$  \\
\includegraphics[width=31mm]{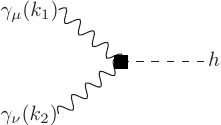} & 0 & $\frac{2g^2\sw^2}{\vev}a_{h\gamma\gamma}\left(k_1\cdot k_2g^{\mu\nu}-k_2^\mu k_1^\nu\right)$ & 0 \\
\includegraphics[width=31mm]{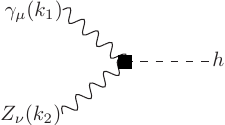} & 0 & $\frac{2g^2\sw\cw}{\vev}a_{h\gamma Z}\left(k_1\cdot k_2g^{\mu\nu}-k_2^\mu k_1^\nu\right)$ & 0 \\
\hline
\includegraphics[width=31mm]{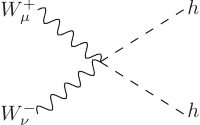} & $i\frac{g^2}{2}g^{\mu\nu}$ & $i\frac{g^2}{2}bg^{\mu\nu}$ & $i\frac{g^2}{2}g^{\mu\nu}$ \\
\includegraphics[width=31mm]{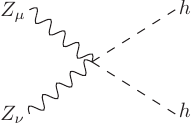} & $i\frac{g^2}{2\cw^2}g^{\mu\nu}$ & $i\frac{g^2}{2\cw^2}bg^{\mu\nu}$ & $i\frac{g^2}{2\cw^2}g^{\mu\nu}$  \\
\hline
\includegraphics[width=31mm]{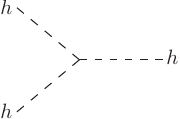} & $-6i\lsm\vev$ & $-6i\lsm\vev\kappa_3$ & $-6i\lahhh\vev$  \\
\includegraphics[width=31mm]{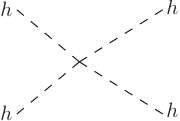} & $-6i\lsm$ & $-6i\lsm\kappa_4$ & $-6i\lahhhh$  \\
\end{tabular}
\caption{Relevant Feynman rules involving the SM-like Higgs boson in the SM, HEFT and 2HDM for comparison. All momenta are incoming.  The relations among the tree-level EW parameters  are as in the SM:  $\mw=(g v)/2$,  $\mz=\mw/\cw$,  $\mh^2=2 \lambda v^2$.  
}
\label{relevant-FR}
\end{table}

\section{One-loop functions}
\label{App-floops}

The  \1loop computation is performed  with dimensional regularization in $D=4-\epsilon$ dimensions and we use the standard definitions for the associated divergence:
\be
\Delta_\epsilon=\frac{1}{4-D}-\gamma_E+\log(4\pi) \,,
\label{div-definition}
\ee
where $\mu_0$ is the usual scale. We implement the compact notation for the momentum integral given by:
\be
\int_k=\mu_0^{4-D}\int\frac{d^D k}{(2\pi)^D} \,.
\ee
To display some results, we also use the scalar two and three-point \1loop integral functions in the Passarino-Veltman notation~\cite{Passarino:1978jh}, with the following conventions:
\bear
  \frac{i}{16\pi^2} A_0(m_1) &=& \int_k \frac{1}{[k^2 - m_1^2]}\,,  \nn\\
  \frac{i}{16\pi^2} B_0(q_1,m_1,m_2) &=& \int_k \frac{1}{[k^2 - m_1^2][(k+q_1)^2 - m_2^2]}\,,  \nn\\
   \frac{i}{16\pi^2}C_0(q_1, q_2, m_1, m_2, m_3) &=& \int_k \frac{1}{[k^2 - m_1^2][(k + q_1)^2 - m_2^2][(k + q_1 + q_2)^2 - m_3^2]}.
\eear
We introduce
\be
f(r)=-\frac{1}{4}\log ^2\left(-\frac{1-\sqrt{1-r}}{1+\sqrt{1-r}}\right) = \left\lbrace
\begin{array}{ll}
\arcsin^2\left(\frac{1}{\sqrt{r}} \right) & \quad r\geq 1 \,, \\
-\frac{1}{4}\left( \ln\left(\frac{1+\sqrt{1-r}}{1-\sqrt{1-r}}\right) -i\pi \right)^2 & \quad 0<r<1 \,,
\end{array}
\right.
\label{fpaper}
\ee
and 
\be
g(r) = -\frac{1}{2}\sqrt{1-r}\log\left(-\frac{1-\sqrt{1-r}}{1+\sqrt{1-r}}\right) = \left\lbrace
\begin{array}{ll}
\sqrt{r-1}\arcsin\left(\frac{1}{\sqrt{r}} \right) & \quad r\geq 1\,,\\
\frac{1}{2}\sqrt{1-r}\left( \ln\left(\frac{1+\sqrt{1-r}}{1-\sqrt{1-r}}\right) -i\pi \right) & \quad 0<r<1 \, ,
\end{array}
\right.
\label{gpaper}
\ee
with the relations
\bear
B_0(q,M,M)&=&\Delta_\epsilon+\log\left(\frac{\mu_0^2}{M^2}\right)+2-2g\left(\frac{4M^2}{q^2}\right) \,,  \nn\\
C_0(0,q,M,M,M)&=&-\frac{2}{q^2}f\left(\frac{4M^2}{q^2}\right)\,.
\label{fg_as_floops}
\eear
The relevant regime for the matching in \eqref{amploop-2HDM-matching} is large $r$, in which case:
\bear
f(r)\vert_{r>>1} &\sim& \frac{1}{r}+\frac{1}{3r^2}+\mO(r^{-3}) \,, \nn\\
g(r)\vert_{r>>1} &\sim& 1-\frac{1}{3r}-\frac{2}{15r^2}+\mO(r^{-3}) \,.
\label{fyg-larger}
\eear

\bibliography{ADHM-matching}

\end{document}